\documentclass{article}
\usepackage{hyperref}
\usepackage[english]{babel}
\usepackage{amssymb,amsmath, amsthm}
\usepackage[all]{xy}

\usepackage{pstricks,pst-node,pst-text,pst-3d}
\usepackage{graphicx}
\usepackage{psfrag}
\newtheorem{thm}{Theorem}[section]
\newtheorem{lemma}{Lemma}[section]
\newtheorem{cor}{Corollary}[section]
\newtheorem{prop}{Proposition}[section]
\newtheorem{defn}{Definition}[section]

\theoremstyle{definition}
\newtheorem{rem}{Remark}[section]

\newcommand{\tr}{\operatorname{tr}\,}

\newcommand{\diag}{\operatorname{diag}}
\newcommand{\etres}{\mathbf{e_3}}

\newcommand{\PP}{\mathcal{P}}  
\newcommand{\w}{\omega}        
\newcommand{\ed}{\mathbf{d}}   
\newcommand{\J}{\mathbf{J}}    
\newcommand{\D}{\mathbf{D}}    

\newcommand{\R}{\mathbb{R}}    
\newcommand{\N}{\mathbb{N}}    
\newcommand{\Ad}{\mathrm{Ad}}  
\newcommand{\ad}{\mathrm{ad}}  
\newcommand{\Sl}{\mathbf{S}}   
\newcommand{\g}{\mathfrak{g}}
\newcommand{\q}{\mathfrak{q}}
\newcommand{\be}{\begin{equation}}
\newcommand{\ee}{\end{equation}}
\newcommand{\bea}{\begin{eqnarray}}
\newcommand{\eea}{\end{eqnarray}}
\newcommand{\II}{\mathbb{I}} 
\newcommand{\rII}{\widehat{\mathbb{I}}_0} 
\newcommand{\Proj}{\mathbb{P}} 
\newcommand{\corr}{\mathrm{corr}\,} 
\newcommand{\Ar}{\mathrm{Ar}} 
\newcommand{\SL}{\mathrm{SL}(3)} 

\newcommand{\Z}{\mathbb{Z}}

\newcommand{\restr}[1]{\vrule height3ex width.4pt depth1.4ex\lower1.4ex\hbox{\scriptsize $\,#1$}}
\newcommand{\rrestr}[1]{\vrule height2ex width.4pt depth0.9ex\lower0.9ex\hbox{\scriptsize $\,#1$}}

\begin{document}

\title{Nonlinear Stability of Riemann Ellipsoids with Symmetric Configurations}

\date{}

\author{Miguel Rodr\'{\i}guez-Olmos\thanks{Ecole Polytechnique F\'ed\'erale de Lausanne (EPFL),
Section de Math\'ematiques. CH-1015 Lausanne, Switzerland, miguel.rodriguez@epfl.ch}, M. Esmeralda Sousa-Dias
\thanks{
Dep.
Matem\'atica, Instituto Superior T\'ecnico, Av. Rovisco Pais,
1049--001 Lisboa, Portugal, edias@math.ist.utl.pt}}

\maketitle

\begin{abstract} We apply geometric techniques to obtain the necessary and sufficient conditions on the existence and nonlinear stability of self-gravitating Riemann ellipsoids having at least two equal axes.
\end{abstract}

\section{Introduction}
A Riemann ellipsoid is a relative equilibrium  for a dynamical model
of a rotating self-gravitating fluid mass that remains an ellipsoid
at all times, and for which  the fluid velocity is a  linear
function of the coordinates. This model, first studied and
formulated by Dirichlet, is nowadays known as Dirichlet's model. The
linearity assumption  on the allowed motions makes the study of
these deformable bodies   very attractive since it implies that
their dynamics is governed by a system of ordinary differential
equations with a finite number of degrees of freedom. These bodies
are also known as affine-rigid bodies or pseudo-rigid bodies.
Dirichlet's model can be viewed as a first order model for the study
of the shape of the Earth, and the study of the stability of its
solutions can be used in planetary stability research.

The study of self-gravitating fluid masses has a  long history which
can be traced back to Newton's times with many contributions from
Dirichlet, MacLaurin, Jacobi, Dedekind, Riemann, Liapunov,
Poincar\'e and Cartan, just to name a few. We can distinguish two
classical approaches to the study of the stability of Riemann
ellipsoids. One  initiated independently in the  latter part  of the
nineteenth century   by Poincar\'e \cite{Poincare} and Liapunov
\cite{Liapunov}  who used expansions in spherical and ellipsoidal
harmonics  to study the stability of MacLaurin ellipsoids and Jacobi
ellipsoids,   not only under Dirichlet's assumptions but also under
more general conditions (allowing perturbations not preserving the
ellipsoidal shape). The other approach occurred in the middle of the
twentieth century with the works of Chandrasekhar and collaborators
who developed   the so-called virial method  by applying it to the
study of the linear stability of Riemann ellipsoids. These works are
collected  in the book of Chandrasekhar \cite{Chand}  which
constitutes a  comprehensive survey on the subject and related
problems with many historical facts on this model.

In recent times the subject has had the attention of several
researchers, in particular in what respects to the application of
new formulations and methods (\cite{Ros88b,Ros01})  to study
rotating deformable bodies, not only subject to the self-gravitating
potential but also for other potentials modelling nuclei (see for
instance \cite{Ros88a}), or elastic bodies (see \cite{CoMu} and
\cite{DeSim}).

Our aim is to use geometric methods not only to obtain a complete
cha\-rac\-terization of the conditions for the existence
 of Riemann ellipsoids having   configurations  with at least two equal axes
 but also  to obtain the complete description
of their nonlinear stability. These geometric methods exploit the
geometry and symmetries of the problem and its Hamiltonian
structure. Some works using the same philosophical approach to
Dirichlet's model  are available in the literature  for studying
several aspects of the problem such as in  \cite{MEres} which
obtains Riemann's theorem as a consequence of the symmetry alone, or
\cite{DebraBif}  where  some results on the formal stability and
bifurcations of MacLaurin spheroids
  are obtained.

  We do not address
the problem of the stability  for Riemann ellipsoids with three
distinct axes for which the self-gravitating potential is an
elliptic integral. In \cite{FaDe} this  has been studied  employing
numerical analysis techniques.

We view Dirichlet's model as a Hamiltonian system where Hamilton's
function, $h$,  has    the form kinetic plus potential energy and is
defined on the cotangent bundle, $T^*\SL$, of the set of all
$3\times 3$ matrices of determinant 1. Furthermore, $h$ is invariant
for the action of
$G=\mathbb{Z}_2\ltimes(\mathrm{SO}(3)\times\mathrm{SO}(3))$ on the
phase space. The $\mathrm{SO}(3)\times\mathrm{SO}(3)$  symmetry
reflects the existence of two conserved vector quantities (by
Noether's theorem): the angular momentum and circulation. The
$\mathbb{Z}_2$ symmetry reflects the reciprocity theorem of
Dedekind: interchanging the angular velocity and vorticity vectors
one obtains another (physically different) solution for Dirichlet's
model with the same geometric configuration.

It is well known that Noether's conserved quantities are organized
as the components of the momentum map. In the case of relative
equilibria with configurations having at least two equal axes
(symmetric configurations)  the corresponding momentum map value can
be singular. Recently, one of the authors developed in
\cite{miguelstab} a method appropriated for the study of nonlinear
stability of this kind of  relative equilibria. This construction
extends the so-called reduced energy-momentum method of Simo, Lewis
and Marsden \cite{Simo} to the case of singular relative equilibria
and so we will refer to the method \cite{miguelstab} as the singular
reduced energy-momentum method. This is the approach we use in this
work to study
 the nonlinear stability of Riemann ellipsoids with
symmetric configurations.

The preliminary sections of the paper are organized as follow: in
Section~\ref{sec geometric formulation} we give the geometric
formulation of Dirichlet's model, in Section~\ref{sec singular
reduced}   the singular reduced energy-momentum method is   briefly
reviewed and in Section \ref{sec potential} we compute the augmented
potential energy for symmetric configurations, as a necessary step
towards the stability analysis.

The main results of the paper are in the following two sections. In
Section~\ref{sec existence}, Theorem~\ref{theorem existence},  we
give the complete characterization of all the possible Riemann
ellipsoids with symmetric configurations.  We prove that  for
Dirichlet's model the only relative equilibria with configurations
having at least two equal axes  are:  the spherical configuration
which has zero angular momentum and circulation, the {\em MacLaurin
spheroids} which are oblate spheroids rotating around  the
(shortest) symmetry axis and have angular momentum and circulation
aligned with it,  and the {\em transversal spheroids}, which have
prolate spheroidal configurations that rotate around an axis, say
$\mathbf{n}$, perpendicular to the (longest) symmetry axis and have
angular momentum and circulation aligned with $\mathbf{n}$. We also
prove that there are no symmetric relative equilibria for which the
angular velocity and vorticity are not aligned with a principal axis
of the relative equilibrium configuration. That  is, there are not
symmetric configurations which are not of type S in Chandrasekhar's
terminology.

In section~\ref{sec stability} we apply the singular reduced
energy-momentum method to the study of the nonlinear stability  of
the relative equilibria found in the previous section. The
main results of this section are theorems~\ref{spherical stability},
\ref{MacLaurin stability} and \ref{transversal stability} giving
respectively necessary and sufficient conditions for the nonlinear stability of the spherical equilibrium,   MacLaurin
spheroids and  transversal spheroids.

Our results on  the eccentricity range for the nonlinear $G_\mu$-stability
 of the MacLaurin spheroids agree with those
already obtained by Liapunov and Poincar\'e (see Remark \ref{rem
maclaurin}). In the works of these authors there is no reference to
the transversal spheroids, however their existence is acknowledged  in
pag.~143 of Chandrasekhar's book \cite{Chand}. It is not clear for
us what are the results obtained by Chandrasekhar with respect to
the stability of the transversal spheroids, however his use of the
virial method only gives linear stability. To the best of our
knowledge, our result on the nonlinear stability of the transversal
spheroids is new.

In conclusion,  this work  presents,  from a purely geometric point
of view,  a self-contained and complete study of the nonlinear
stability of all sym\-me\-tric relative equilibria  for  Dirichlet's
model. At the same time,  the richness of the model helps    to
clarify the applicability of the singular reduced energy-momentum
method, providing also a methodology for its application to other
models like.

\noindent {\bf Acknowledgements.} We would like to thank to Centro
de An\'alise Matem\'atica, Geometria e Sistemas Din\^amicos of the
IST, Lisbon, for the portuguese translation of  original Riemann's
 paper \cite{Riemann} made by C.E. Harle. The work of ESD has been
supported by the Funda\c{c}\~ao para a Ci\^encia e a Tecnologia
through the Program POCI 2010/FEDER.

\section{Geometric formulation of Dirichlet's model}\label{sec geometric formulation}

Let $(M,\ll\cdot,\cdot\gg)$ be a Riemannian manifold (the
configuration manifold), $G$ a Lie group that acts by isometries on
$M$ (the symmetry group) and $V\in C^\infty (M)$ a $G$-invariant
function (the potential energy). With these ingredients we construct
a symmetric Hamiltonian system on $T^*M$ (which is a manifold
equipped with a natural symplectic structure) as follows: the
potential energy $V$ can be lifted to $T^*M$ with the pullback of
the cotangent bundle projection $\tau:T^*M\rightarrow M$. We denote
this lifted function also by $V$. The Riemannian metric on $M$
induces an inner product on each cotangent fiber $T^*_xM$, $x\in M$.
Then the Hamiltonian is defined as
\begin{equation*}h(p_x)=\frac 12 \|p_x\|^2+V(x),\quad p_x\in T_x^*M.\end{equation*} The
$G$-action on $M$ induces a cotangent-lifted Hamiltonian action on
$T^*M$ with associated equivariant momentum map $\J:T^*M\rightarrow
\g^*$ defined by
\begin{equation*}\langle \J(p_x),\xi\rangle=\langle
p_x,\xi_M(x)\rangle\quad\forall\,\xi\in\g,\end{equation*} where $\xi_M$
is the fundamental vector field on $M$ associated to the generator
$\xi$, defined by
$$\xi_M(x)=\frac{d}{dt}\restr{t=0}e^{t\xi}\cdot x.$$
This momentum map is $\text{Ad}^*$-equivariant in the sense that
$\J(g\cdot p_x)=\Ad_{g^{-1}}^*\J(p_x)$ for every $p_x\in T_x^*M$,
$g\in G$.

 The
Hamiltonian $h$ is $G$-invariant for this lifted action (this
follows from the invariance of the metric and of $V$). Therefore,
due to Noether's theorem, the components of $\J$ are conserved
quantities for the Hamiltonian dynamics associated to $h$. The
quadruple $(M,\ll\cdot,\cdot\gg,G,V)$ is called a \emph{symmetric
simple mechanical system}.

Let $(M,\ll\cdot,\cdot\gg,G,V)$ be a simple mechanical system with
symmetry. A \emph{relative equilibrium} is a point in phase space
$p_x\in T^*M$ such that its Hamiltonian orbit lies inside a group
orbit for the cotangent-lifted action. This amounts to the existence
of a generator $\xi\in\g$ such that the evolution of $p_x$ is given
by $e^{t\xi}\cdot p_x$. The element $\xi$ is called a
\emph{velocity} for the relative equilibrium and is defined up to
addition of elements in $\g_{p_x}=\text{Lie}\,(G_{p_x})$, where
$G_{p_x}$ is the stabilizer of $p_x$ under the cotangent-lifted
action. A useful criterion for finding relative equilibria in simple
mechanical systems is given by the following theorem:
\begin{thm}[Marsden \cite{MarsdenLN}]  \label{REthm}A point  $p_x\in T^*M$ of a symmetric simple mechanical system $(M,\ll\cdot,\cdot\gg,G,V)$ is a relative equilibrium  with  velocity $\xi\in\g$
if and only if the following conditions are verified:
\begin{enumerate}
\item $p_x = \ll \xi_M(x) , \cdot\gg$.
\item  $x$ is a critical point of $V_\xi:=V-\frac 12\langle\xi,\II(\cdot)(\xi)\rangle$,
\end{enumerate}
where $\II:M\times\g\rightarrow\g^*$ is defined by
$\langle\xi,\II(x)(\eta)\rangle=\ll\xi_M(x),\eta_M(x)\gg$.
Moreover, the momentum $\mu=\J(p_x)\in\g^*$ of the relative
equilibrium is given by $\mu = \II(x)
(\xi). $
\end{thm}
Note that, in virtue of the above theorem, any relative equilibrium
is characterized by a configuration-velocity pair $(x,\xi)\in
M\times\g$ satisfying $\ed V_\xi(x)=0$. The map $\II$ is called the
\emph{locked inertia tensor}, while the function $V_\xi$ is called
the \emph{augmented potential}. We indicate for later use that the
kernel of $\II$ is precisely $\g_x$, the Lie algebra of $G_x$, the
stabilizer of $x$ for the $G$-action on $M$. The knowledge of the
pair $(x,\xi)$ allows us to compute the stabilizer of the
corresponding relative equilibrium $p_x=\ll\xi_M(x),\cdot\gg$ with
the formula (see \cite{RSD}):
\begin{equation*}G_{p_x}=\{g\in G_x\,:\,\Ad_g\xi-\xi\in\g_x\}.\end{equation*}

Dirichlet's model is a symmetric simple mechanical system for the motion of
 a homogenous and incompressible fluid mass of density
$\rho$ having as reference configuration the unit ball centered at
the origin in $\R^3$ and subject to the self-gravitating potential.
The only allowed configurations for this model are linear embeddings
of the reference ball into $\R^3$ preserving volume and orientation.
The configuration manifold $M$ for a self-gravitating fluid mass
under Dirichlet's conditions is then  $\mathrm{SL}(3)$, the group of
all $3\times 3$ matrices with determinant equal to 1, which is
equivalent to the space of orientation and volume preserving linear
automorphisms of $\R^3$. In what follows we review the geometric
formulation of Dirichlet's model as a symmetric simple mechanical
system on $GL^+(3)$, the group of all $3\times 3$ matrices with
positive determinant, with a symmetric holonomic constraint.

The singular value decomposition of any linear automorphism of
$\R^3$ allows to decompose (non-uniquely)  any matrix
$F\in\mathrm{GL}^+(3)$ as
\begin{equation*}
F = L AR^T
\end{equation*}
where $L,R\in \mathrm{SO}(3)$, and $A$ is a diagonal matrix  with
positive entries called singular values (the square roots of the
eigenvalues of $C=F^TF$). It follows from this decomposition that
the reference unit ball is mapped by $F$ into a solid ellipsoid
 of
equation $\mathbf{X}\cdot C^{-1}\mathbf{X}=1,\,\mathbf{X}\in\R^3$, having
 principal axes half-lengths equal  to the entries  of $A$.
Physically  the matrix $L$ describes the rigid rotation of  the body
in space relative to an inertial frame and  $R$ is related to the
rigid internal motion of the fluid with respect to a moving frame.
Then $A$ is an orientation preserving dilatation of the original
reference body into an ellipsoid with principal axes aligned with
the eigenvectors of $A$. The condition on the volume preservation of
the total embedding corresponds to impose the holonomic constraint
$\det F=1$ (or equivalently $\det A=1$), which in turn amounts to consider
our system as defined on $\mathrm{SL}(3)$.

 The tangent space at $F\in \mathrm{GL}^+(3)$ is isomorphic to $\mathrm{L}(3)$, the vector space of $3\times 3$ matrices.
We can define a Riemannian metric on $\mathrm{GL}^+(3)$ as:
\begin{equation}\label{riem}
\ll \delta F_1,\delta F_2\gg =T\,  \operatorname{tr} (\delta
F_1^T\delta F_2)\end{equation}  for $\delta F_1,\delta F_2\in
T_F\mathrm{GL}^+(3)$, and $T$ is a constant depending on the density
of the reference body and other physical parameters of the system.
In the case of interest here, the reference body is a homogeneous
unit ball of constant density $\rho$, and  $T$ in  \eqref{riem} is
\begin{equation*} T=\frac{4\pi}{15}\rho.\end{equation*}

The symmetry group $G$ of our model is the  semi-direct product
$G=\mathbb{Z}_2\ltimes(\mathrm{SO}(3)\times\mathrm{SO}(3))$, where
$\mathbb{Z}_2=\{e,\sigma\}$. Several actions of $G$ of interest in this paper are:
\begin{enumerate}
\item[(1)] The $G$-action on $G$: If $(\gamma;g,h),(\gamma';g',h')\in \mathbb{Z}_2\ltimes(\mathrm{SO}(3)\times\mathrm{SO}(3))$ then
\begin{equation*} (\gamma;g,h)\cdot
(\gamma';g',h')=(\gamma\gamma';(g,h)\cdot(\gamma\cdot
(g',h')),\end{equation*}
 where for the nontrivial element $\sigma\in\mathbb{Z}_2$, $\sigma\cdot (g',h')=(h',g')$, and
$\mathrm{SO}(3)\times\mathrm{SO}(3)$ acts on itself by the direct product of left matrix multiplications.

\item[(2)] The adjoint representation of $G$:
We identify the Lie algebra of $SO(3)$, the set  $\mathfrak{so}(3)$ of skew-symmetric
$3\times 3$ matrices, with $\R^3$ via the usual
isomorphism  $\,\widehat{ }:\R^3\rightarrow \mathfrak{so}(3)$:
\begin{equation}
{\boldmath v} = (v_{1}, v_{2}, v_{3}) \mapsto {\widehat {\boldmath
v}} =
\begin{bmatrix}\label{skewsymm}
                   0 & -v_{3} & v_{2} \\
                   v_{3} & 0 & -v_{1} \\
                   -v_{2} & v_{1} & 0 \end{bmatrix}
\end{equation}
This  is an isomorphism of Lie algebras, i.e $(\mathfrak{so}(3), [\cdot\,
, \, \cdot])$ is isomorphic by \eqref{skewsymm} to $(\R^3, \times)$, where $ [\, , \, ]$
denotes the commutator of matrices and $\times$ denotes the vector
product of  vectors in $\R^3$.

The Lie algebra of $G$ is then $\g=\mathbb{R}^3\oplus\mathbb{R}^3$
and the adjoint action  is given by
\begin{equation*}
\Ad_{(\gamma ; g,h)}
(\xi_L,\xi_R)=\gamma\cdot(g\cdot\xi_L,h\cdot\xi_R)
\end{equation*}
where $\sigma\cdot (\xi_L,\xi_R)=(\xi_R,\xi_L)$ and $g\cdot\xi_L$ is
the rotation of $\xi_L$ by $g$ (and similarly for $\xi_R$).

Using the standard inner product  in $\R^3$ we also identify $\g^*$
with $\mathbb{R}^3\oplus\mathbb{R}^3$. Under this identification it follows easily that the coadjoint representation has the expression
\begin{equation*}
\Ad^*_{(\gamma ; g,h)^{-1}}
(\mu_L,\mu_R)=\gamma\cdot(g\cdot\mu_L,h\cdot\mu_R)
\end{equation*}

\item[(3)] The $G$-action on $\mathrm{GL}^+(3)$:
$$(e;L,R)\cdot F  =  LFR^T,\quad
(\sigma;L,R)\cdot F  =  RF^TL^T,$$
\end{enumerate}
Note that the $\mathbb{Z}_2$ transposition  symmetry on
$\mathrm{SL}(3)$ (first noticed by Dedekind) maps a rigidly rotating
configuration without internal motion  into one that is stationary
in space but with rigid fluid internal motions. That is,  for a given
ellipsoid there is an adjoint one, obtained by transposition.
These adjoint type of ellipsoids are called Dedekind ellipsoids. 
More generally, the transposition symmetry interchanges external
rotations and internal motions for any solution of Dirichlet's
model.

Any $G$-invariant function $f$ on $\mathrm{GL}^+(3)$ can be written
as $$f(F)=\tilde{f}(I_1(F),I_2(F),I_3(F)),$$ where
$\tilde{f}:\R^3\rightarrow \R$, and $I_1, I_2$ and $I_3$ are the
three principal invariants of a $3\times 3$ matrix, given by
\begin{eqnarray}\label{I1} I_1(F) & = & \mathrm{tr}(S)\\
\label{I2} I_2(F) & = &\frac 12 (\mathrm{tr}^2(S)-\mathrm{tr}(S^2))\\
 I_3(F)&=&\det (S)  \label{I3}
\end{eqnarray} with
$S=FF^T$ (also valid interchanging $S$ with $C=F^TF$). Note that
$I_1,I_2,I_3$ are $G$-invariant and that $1$ is a regular value of
$I_3$. Hence we have $\mathrm{SL}(3)=I_3^{-1}(1)$ as a $G$-invariant
submanifold of $\mathrm{GL}^+(3)$. The condition $I_3=1$ is the
holonomic constraint of the model. Note also that the restriction of
a $G$-invariant function $f\in C^G(\mathrm{GL}^{+}(3))$ to $\SL$ is given by
$$f(F)=\tilde{f}(I_1(F),I_2(F),1),$$
for $F\in\mathrm{SL}(3)\subset \mathrm{GL}^+(3)$. Therefore any
$G$-invariant function on $\SL$ can be written as
$h(F)=\widehat{h}(I_1(F),I_2(F))$, with $\widehat{h}:\R^2\rightarrow
\R$. Any $G$-invariant function on $\mathrm{SL}(3)$ can be then
extended invariantly to $\mathrm{GL}^+(3)$ by declaring it to be
independent of $I_3(F)$. From now on we will drop the tildes and
hats from the corresponding functions unless there is risk of
confusion.

Since $G$ acts on $\mathrm{GL}^+(3)$ by isometries with respect to
\eqref{riem}, the induced metric on $\mathrm{SL}(3)$ (which we will
denote by the same symbol) is also $G$-invariant. For later use we recall
that tangent vectors to $\mathrm{SL}(3)$ can be seen as tangent
vectors to $\mathrm{GL}^+(3)$ satisfying the linearization of the
constraint $I_3=1$. In other words,
$$T_F\mathrm{SL}(3)=\{\delta F\in \mathrm{L}(3)\, :\,\operatorname{tr} (F^{-1}\delta
F)=0\}.$$

Dirichlet's model is governed by a Hamiltonian function of the
form kinetic  plus potential energy on the phase space  $\PP =
T^*\mathrm{SL}(3)$ given by
\begin{equation*}h(p_F)=\frac T2 \|p_F\|^2+V(F),\quad p_F\in T_F^*\mathrm{SL}(3).\end{equation*} Here
$\|p_F\|$ is the norm of the covector $p_F$ (seen as a $3\times 3$
matrix) relative to the metric on $\mathrm{SL}(3)$. The
potential energy $V$ for a self-gravitating body of homogeneous density $\rho$ under Dirichlet's
assumptions is given by restricting the function
\begin{equation}\label{gravity}
V(F)=-R\int_0^\infty \frac{ds}{\Delta(F)},\end{equation} where
$F\in\mathrm{SL}(3),\,R=\frac{8}{15}\pi^2 G\rho^2$, $G$ is the gravitational constant and
\begin{equation}\label{Delta}
\Delta (F)=\sqrt{s^3+I_1(F)s^2+I_2(F)s+1}.
\end{equation}
The quadruple $(\mathrm{SL}(3),\ll\cdot,\cdot\gg,\mathbb{Z}_2\ltimes
(\mathrm{SO}(3)\times\mathrm{SO}(3)),V)$ defines a symmetric simple
mechanical system on $\mathrm{SL}(3)$.

The infinitesimal generator for the $G$-action on
$M=\mathrm{GL}^+(3)$ (and on $\SL$) corresponding  to $\xi = (\xi_L,
\xi_R)\in\R^3\times \R^3$ is:
\begin{equation}\label{infgene}
\begin{matrix}
\xi_M(F) &= \frac{d}{dt}\rrestr{t=0} (\exp t\widehat{\xi_L}, \exp
t\widehat{\xi_R})\cdot F= \widehat{\xi_L} F - F \widehat{\xi_R}.
\end{matrix}
\end{equation}
The vectors $\xi_L$ and $\xi_R$ are respectively the angular
velocity and vorticity of the fluid motion. We denote the momentum
value of $p_F$ by $\J(p_F) = (\mathfrak{j}, \mathfrak{c})$. The
components $\mathfrak{j}$ and $\mathfrak{c}$ are respectively  the
angular momentum and circulation of the instantaneous state $p_F$
(see for instance \cite{MEres} or \cite{Chand}).

A Riemann ellipsoid (a.k.a. an ellipsoidal figure of equilibrium) is
a solution of the Hamiltonian system  defined by Dirichlet's model
with angular velocity, vorticity and principal axes lengths all
constant. In our setting, Riemman ellipsoids correspond exactly to
relative equilibria of the underlying symmetric simple mechanical
system. Therefore, a Riemann ellipsoid is represented by a triple
$(F,\xi_L,\xi_R)$, where $F\in \mathrm{SL}(3)$ is the configuration
matrix and the Lie algebra element $(\xi_L,\xi_R)\in \g$ is the
angular velocity-vorticity pair.

\section{The singular reduced energy-momentum method}\label{sec singular reduced}

In the last years, several are the works studying the stability of
relative equilibria of Hamiltonian  systems by exploiting their
symmetry and the geometric properties of their phase space (see for
instance \cite{Arnold2}, \cite{Patrick92}, and \cite{MarsdenLN} for
a overview). Most of these methods can be used to test the stability
of relative equilibria lying in singular level sets of the momentum
map, for instance \cite{LermanSinger} and \cite{OrtegaRatiu} under
the hypotesis of  $G_\mu$ compact and  \cite{Montaldi},
\cite{PaRoWu} using topologic properties. This observation is
important since, as we will see, the class of symmetric Riemann
ellipsoids known as MacLaurin spheroids corresponds precisely to
non-trivial relative equilibria for Dirichlet's model having
singular momentum values. We refer the reader to \cite{PaRoWu}  for
a  comparison of the applicability of several existing methods.

The  generally adopted notion of stability for relative equilibria
of symmetric Hamiltonian systems  is that of $G_\mu$-stability,
introduced in \cite{Patrick92} and that  we now review in the
context of symmetric simple mechanical systems. This notion  is
closely related to the Liapunov stability of the induced Hamiltonian
system on the reduced phase space.
\begin{defn}\label{patrick} Let $(M,\ll\cdot,\cdot\gg,G,V)$ be a symmetric simple mechanical system and $p_x\in T^*M$ a
relative equilibrium  with momentum value $\mu=\J(p_x)$. We say that
$p_x$ is $G_\mu$ stable if for every $G_\mu$-invariant neighborhood
$U\subset T^*M$ of the orbit $G_\mu\cdot p_x$ there exists a
neighborhood $O$ of $p_x$ such that the Hamiltonian evolution of $O$
lies in $U$ for all time.
\end{defn}
In the early 1990's (\cite{Simo}) a tool known as the reduced
energy-momentum method  has been developed, providing sufficient
conditions for the stability of relative  equilibria of a simple
mechanical system under the hypothesis  that its momentum is a
regular value of the momentum map. This method is especially
well-suited for simple mechanical systems since in incorporates all
of their distinguishing  characteristics with respect to general
Hamiltonian systems. Recently,  based on the characterization
\cite{PerlRodSou} of the so-called symplectic normal space $N$ for a
cotangent-lifted action,  the reduced energy-momentum method was
generalized in \cite{miguelstab} to cover also the case of singular
momentum values.

In this section we outline the implementation of this singular reduced energy-momentum method following \cite{miguelstab}. Our setup will be as in Definition \ref{patrick} and Section 2. In particular we will fix a relative equilibrium $p_x$ with configuration-velocity pair $(x,\xi)$ and momentum $\mu$. We will also assume that the $G$-action on $M$ is proper and that there exists a $G_\mu$-invariant complement to $\g_\mu$ in $\g$. These last two conditions are always satisfied for any relative equilibrium in Dirichlet's model due to the compactness of $G$.

We start by stating some key observations: First, by equivariance of
$\J:T^*M\rightarrow \g^*$ and $\tau:T^*M\rightarrow M$, one has
$G_{p_x}\subset G_x$ and $G_{p_x}\subset G_\mu$. In fact, it is not
difficult to prove the characterization
\begin{equation}\label{momentumcarac}G_{p_x}=G_x\cap G_\mu.\end{equation}
We remark that the above formula is not valid in general for covectors $p_x$ other than relative equilibria.

Second, also by equivariance of $\tau$ together with the Bifurcation
Lemma (see \cite{AMM81}),  if $\mu$ is a singular momentum value
then $\g_x\neq \{0\}$, in which case $\mu\in(\g_x)^\circ$. Third,
the properness of the $G$-action implies that $G_{p_x}$ is compact.
This, together with \eqref{momentumcarac}  allows to define the
following $G_{p_x}$-invariant splittings:
\begin{equation}\label{split}
\g_\mu =\g_{p_x}\oplus \mathfrak{p}\qquad\text{and}\qquad \g = \g_x\oplus\mathfrak{p}\oplus \mathfrak{t},
\end{equation} which by duality induce the splittings
\begin{equation}\label{dualsplit}
\g_\mu^* =\g_{p_x}^*\oplus \mathfrak{p}^*\qquad\text{and}\qquad \g^* = \g_x^*\oplus\mathfrak{p}^*\oplus \mathfrak{t}^*.
\end{equation}
Here $\mathfrak{p}$ and $\mathfrak{t}$ must be chosen in such a way
that $\II (x)(\mathfrak{p})\subset\mathfrak{t}^\circ$.  Note, from
the definition of the locked inertia tensor in Theorem \ref{REthm},
that $\ker \II(x)=\g_x$, so the restriction
\begin{equation*}\rII=\II(x)\rrestr{\mathfrak{p}\oplus\mathfrak{t}}:\mathfrak{p}\oplus\mathfrak{t}\rightarrow
(\g_x)^\circ = \mathfrak{p}^*\oplus\mathfrak{t}^*\end{equation*} is a
$G_{p_x}$-equivariant isomorphism. Then the condition on the above
splitting is that $\mathfrak{p}$ and $\mathfrak{t}$ must be
orthogonal with respect to the inner product on
$\mathfrak{p}\oplus\mathfrak{t}$ induced by $\rII$.

We will denote generically the linear projections associated to the
splittings \eqref{split} and \eqref{dualsplit} by the letter $\Proj$
with an appropriate subindex. For instance
$\Proj_\mathfrak{p}:\g\rightarrow\mathfrak{p}$ or
$\Proj_{\mathfrak{t}^*}:\g^*\rightarrow\mathfrak{t}^*$. It is a
consequence of Noether's theorem that  $\xi\in\g_\mu$ and  so we
will denote by  $\xi^\perp=\Proj_\mathfrak{p}(\xi)\in \mathfrak{p}$
(the  orthogonal velocity of the relative equilibrium).

In this work we only need a particular version of the singular reduced energy-momentum method. For,
 consider the following definitions:
\begin{itemize}
\item[(1)] Let $\Sl$  be the linear orthogonal slice for the $G$-action on $M$ at $x$, i.e. $$\Sl=(T_x(G\cdot x))^\perp\in T_xM.$$
\item[(2)] Define the subspace $\q^\mu\subset\g$ as
\begin{equation*}
\mathfrak{q}^\mu =\left\{\gamma\in\mathfrak{t}: \,\, \Proj_{\g^*_x}\left[\ad^*_\gamma\mu\right]=0\right\}.
\end{equation*}
\item[(3)] Define the space of internal variations
\begin{equation}\label{sigmaint}
\Sigma_{\text{int}} =
 \left\{\gamma^a_M+a : \,\, \gamma^a\in\mathfrak{q}^\mu,\, a\in\Sl ,\, \,\text{and}\,\, \left(\D\II \cdot \left(\gamma^a_M(F)+a\right)\right) (\xi^\perp)\in\mathfrak{p}^*\right\}.
  \end{equation}
 \item[(4)] For any $v_1,v_2\in T_xM$,  the correction term is  the bilinear form on $T_xM$ defined by
 \begin{equation}\label{corrterm}
\corr_\xi (x) (v_1, v_2) = \left\langle
\Proj_{\mathfrak{p}^*\oplus\mathfrak{t}^*} \left[\left(\D\II\cdot
v_1\right) (\xi)\right]\, ,\,
\rII^{-1}\left(\Proj_{\mathfrak{p}^*\oplus\mathfrak{t}^*}\left[\left(\D\II\cdot
v_2\right) (\xi)\right]\right) \right\rangle.
\end{equation}
\item[(5)] The Arnold form $\Ar: \mathfrak{q}^\mu\times \mathfrak{q}^\mu\rightarrow \R$ is defined by:
 \begin{equation}\label{Arnold form}
\Ar\, (\gamma_1, \gamma_2) = \left\langle\ad^*_{\gamma_1}\mu\, ,\,
\rII^{-1}\left(\ad_{\gamma_2}\mu\right)+\Proj_{\mathfrak{p}\oplus\mathfrak{t}}\left[\ad_{\gamma_2}\left(\rII^{-1}\mu\right)\right]
\right\rangle.
\end{equation}
\end{itemize}

The following theorem (Corollary 6.2 of \cite{miguelstab}) is the synthesis of the singular reduced energy-momentum method.
\begin{thm}[]\label{energy miguel} Let $p_x\in T^*M$ be a relative equilibrium  with configuration-velocity pair $(x,\xi)\in M\times \g$ and
momentum $\mu\in\g^*$ such that $\dim (G\cdot F)<\dim M$. Let
$\xi^\perp=\Proj_\mathfrak{p}(\xi)$ be its orthogonal velocity. If the
Arnold form is non-degenerate at $p_x$ and
$\left(\ed^2_xV_{\xi^\perp}+
\corr_{\xi^\perp}(x)\right)\restr{\Sigma_{\mathrm{int}}}$ is
positive definite, then the relative equilibrium is $G_\mu$-stable.
\end{thm}

When the Arnold form is non-degenerate it is also shown in
\cite{miguelstab} that the symplectic matrix of the symplectic
normal space at $p_x$ has a particularly simple block-diagonal
expression. We quote this result which will be essential in the
proof of the linear unstability of some Riemann ellipsoids.

\begin{thm} \label{symplectic mat} If the Arnold form is non-degenerate at the relative equilibrium $p_x$  with configuration-velocity pair $(x,\xi)$ and momentum $\mu$, the symplectic normal space at $p_x$, is symplectomorphic to $N=\q^\mu\oplus\Sigma_\mathrm{int}\oplus\Sl^*$ equipped with the symplectic matrix
\begin{equation*}
\begin{array}{ccccc}
 & & \q^\mu & \Sigma_\mathrm{int} & \Sl^*\\
\w_N & = & \left[\begin{array}{c} \Xi \\ \Psi^T \\ 0 \end{array}\right. & \begin{array}{c} -\Psi \\
-\ed\chi^{\xi^\perp}\rrestr{\Sigma_\mathrm{int}} \\ -\mathbf{1} \end{array} & \left. \begin{array}{c} 0 \\
\mathbf{1}
\\ 0 \end{array}\right]
\end{array}
\end{equation*}
where
\begin{align*}
 \Xi
(\gamma_1,\gamma_2) & =  -\langle
\mu,\ad_{\gamma_1}\gamma_2\rangle,\qquad \qquad
\Psi (\gamma,(\gamma^b_M(x)+b))  =  \langle\mu,\ad_\gamma
\gamma^b\rangle
\end{align*}
and $\chi^{\xi^\perp}$ is the one-form defined by
$\chi^{\xi^\perp}(X)=\ll\xi^\perp_M,X\gg$, for all
$X\in\mathfrak{X}(M)$.
\end{thm}

\begin{rem}\label{orbitRE}
The $G$-invariance of $V$ and $\ll\cdot,\cdot\gg$ imply the
following property. If $(x,\xi)$ is a relative equilibrium, then the
orbit $(g\cdot x,\Ad_g\xi)$ for every $g\in G$ consists of relative
equilibria with the same stability or unstability properties.
\end{rem}

\begin{rem}\label{augmentations}
The reason why in the previous section we looked at Dirichlet's
model as a simple mechanical system holonomicaly constrained is that
the unconstrained space $\mathrm{GL}^+(3)$ is an open domain of the
vector space $\mathrm{L}(3)$, and then the implementation of the
reduced energy-momentum method is easier than if one is working
directly on $\mathrm{SL}(3)$. In view of the survey of the method,
the strategy will be to use the trivial extension of the
self-gravitating potential to $\mathrm{GL}^+(3)$ and consider its
augmented potential with respect to the locked inertia tensor
corresponding to the original Riemannian metric on
$\mathrm{GL}^+(3)$. Then we further augment this augmented potential
with the constraint function $I_3$ and Lagrange multiplier
$\lambda$. Denoting by
$$V^\lambda_{\xi_L,\xi_R}=V_{\xi_L,\xi_R}-\lambda\det$$ the
resulting twice augmented potential, we have:
\begin{enumerate}
\item[(1)] Relative equilibria for Dirichlet's model
correspond to triples $(F,\xi_L,\xi_R)$ with
$F\in\mathrm{GL}^+(3),\,(\xi_L,\xi_R)\in \R^3\times\R^3$ such that
the following two equations hold:
\begin{equation}\label{existence2augmented}\ed
V^\lambda_{\xi_L,\xi_R}(F)=0,\quad\text{and}\quad
\det(F)=1.\end{equation}

\item[(2)] The stability test now follows from the following
observation. If we call $\Sigma_\text{int}^{\mathrm{GL}^+(3)}$ and
$\Sigma_\text{int}^{\mathrm{SL}(3)}$ the spaces of internal
variations for $\mathrm{GL}^+(3)$ and $\mathrm{SL}(3)$ associated to
the triple $(F,\xi_L,\xi_R)$, according to \eqref{sigmaint}, we
notice that
$$\Sigma_\text{int}^{\mathrm{SL}(3)}=\Sigma_\text{int}^{\mathrm{GL}^+(3)}\cap
\ker T_FI_3.$$ Therefore, according to the general method, and
standard Lagrange multiplier theory, to conclude stability it
suffices to study the eigenvalues of the bilinear form
$$\left(\ed^2_FV^\lambda_{(\xi_L,\xi_R)^\perp}+\corr_{(\xi_L,\xi_R)^\perp}(F)\right)\restr{\Sigma_\text{int}^{\mathrm{SL}(3)}},$$
where $\Sigma_\text{int}^{\mathrm{SL}(3)}$ is seen as a vector
subspace of $\Sigma_\text{int}^{\mathrm{GL}^+(3)}$. From now on we
will omit the superindex $\SL$ for the space of internal variations.
\end{enumerate}
\end{rem}

\section{The augmented self-gravitating potential for symmetric configurations}\label{sec potential}

In this section we compute the augmented potential $$V_{\xi_L,\xi_R}
= V-\frac 12\langle(\xi_L,\xi_R),\II(F)(\xi_L,\xi_R)\rangle$$ in the
unconstrained configuration space $\mathrm{GL}^+(3)$ and collect some
results for the self-gravitating potential $V$. The potential $V(F)$
at a typical configuration is an elliptic integral except for
symmetric configurations (i.e. with at least two equal
singular values) for which it can be integrated by elementary functions. The
extension to the unconstrained space of the potential $V$ depends on
$F\in\mathrm{GL}^+(3)$ through the two principal invariants $I_1$
and $I_2$ defined by \eqref{I1} and \eqref{I2} respectively. In the
study of the existence and stability of relative equilibria of
simple mechanical systems will be necessary to compute the first and
second derivatives of $V_{\xi_L,\xi_R}$ and the results of this
section are essential to this end. In the following we will
restrict ourselves to diagonal configurations only. There is no loss
of generality with this assumption since, according to the singular value decomposition,
every matrix $F\in\mathrm{GL}^+(3)$ belongs to the $G$-orbit of some
diagonal configuration $D$ by some element $(e;A,B)\in G$. Hence
by Remark \ref{orbitRE} the qualitative properties of a relative
equilibrium $(D,(\xi_L,\xi_R))$ are the same as those of
$(ADB^T,(A\xi_L,B\xi_R))$.

Let $J(k,r)$, with $k,r\in \N$ be the following family of integrals:
\begin{equation*}
J(k,r):=\int_0^\infty \frac{s^rds}{\Delta (F)^k},\end{equation*} and
denote by $V_i$ ($i=1,2$) the partial derivative of $V$ with respect
to $I_i$ and by $V_{ij}$ the partial derivative of $V_i$ with
respect to $I_j$ for $j=1,2$. Using  \eqref{gravity}, elementary
calculus computations give:
\begin{align}\label{first derivatives}
V & =  -R J(1,0) & \,  V_1 & =  \frac{R}{2} J(3,2) &\, V_2 & =
\frac{R}{2} J(3,1)&\,  \\
\label{second derivatives} V_{11} & =  - \frac {3R}{4}  J(5,4) &\,
V_{12} & =  - \frac {3R}{4}  J(5,3) &\,  V_{22} & =  - \frac {3
R}{4} J(5,2).&
\end{align}
Note that the integrals $J(k,r)$ are all positive as well as $V_1$
and $V_2$.

Next proposition gives the values of $J(k,r)$ in the case of
spheroidal (two equal axes) configurations.

\begin{prop} \label{integrals}Let $F=\operatorname{diag} (a,a,c)$ be a spheroidal configuration with $a^2c=1$.
 \begin{itemize}
\item[(i)]  The integral $J(k,r)$ for the oblate spheroid $F$ ($a>c$)  and  eccentricity $e = \sqrt{1- \left(\frac{c}{a}\right)^2}$ is given by
\begin{equation}\label{JO}
J_O(k,r)=\frac{2}{(1-e^2)^{\frac{2(r+1)-3k}{6}}}\int_0^1\frac{(1-x^2)^rx^{3(k-1)-2r}dx}{(1-e^2x^2)^\frac
k2}.
\end{equation}
\item[(ii)] The integral $J(k,r)$ for the prolate spheroid $F$ ($a< c$) and eccentricity $e = \sqrt{1- \left(\frac{a}{c}\right)^2}$ is given by
\begin{equation}\label{JP}
J_P(k,r)=\frac{2}{(1-e^2)^{\frac{2(r+1)-3k}{3}}}\int_0^1\frac{(1-x^2)^rx^{3(k-1)-2r}dx}{(1-e^2x^2)^{k}}.\end{equation}
\end{itemize}
\end{prop}

\begin{proof} Note that for the diagonal configuration $F=\diag (a,b,c)$ the value of $\Delta (F)$ in definition \eqref{Delta}  is
$$\Delta (F)= [(a^2+s)(b^2+s) (c^2+s)]^{1/2}.$$
For (i): making the change of variables $s= a^2 \tan^2\theta$ we get
\begin{align*}
J_O(k,r)&= 2 a^{2(r+1)-3k} \int_0^{\pi/2}
\frac{\left(\sin\theta\right)^{2r+1} \left(\cos \theta\right)^{3(k-1)-2 r} }{\left(1+\frac{c^2-a^2}{a^2}\cos^2\theta\right)^{k/2}}\, d\theta.
\end{align*}
Since the eccentricity is  $e^2=  \frac{a^2-c^2}{a^2}$ then
$a=(1-e^2)^{-1/6}$ and $c=(1-e^2)^{1/3}$ because $a^2c=1$. Then,
from the above expression  for $J_O(k,r)$ one gets
\begin{align*}
J_O(k,r)&= \frac{2}{(1-e^2)^{\frac{2(r+1)-3k}{6}}} \int_0^{\pi/2}
\frac{\left(\sin\theta\right)^{2r+1} \left(\cos \theta\right)^{3(k-1)-2 r} }{\left(1- e^2\cos^2\theta\right)^{k/2}}\, d\theta.
\end{align*}
Making $x= \cos\theta$ the result follows.

For (ii): The change of variables $s= c^2 \tan^2\theta$ gives
\begin{align*}
J_P(k,r)&= 2 c^{2(r+1)-3k} \int_0^{\pi/2}
\frac{\left(\sin\theta\right)^{2r+1} \left(\cos \theta\right)^{3(k-1)-2 r} }{\left(1+\frac{a^2-c^2}{c^2}\cos^2\theta\right)^{k}}\, d\theta.
\end{align*}
The  eccentricity $e$ of the prolate spheroid is such that  $c^2(1-e^2) = a^2$.  As $a^2c=1$ then   $a=(1-e^2)^{1/6}$ and $c=(1-e^2)^{-1/3}$ and   the result follows for $x= \cos\theta$.
\end{proof}

As stated in the previous section, in order to find critical points
of a $G$-invariant function defined in $\SL$ we will work with its
extension to $\mathrm{GL}^+(3)$ subject to the constraint $\det F=1$. For, since any such function can be written as $f(F) = f(I_1(F),I_2(F))$, in order to compute the critical points we use the augmented function
$f^\lambda(F) = f(I_1(F),I_2(F))-\lambda \det (F)$ subject to the
condition $\det (F)=1$. For the differentiation of $f^\lambda$
consider the pairing between vectors $\delta F\in T_FSL(3)$ and
covectors $B\in T^*_F\SL$ to be $B\cdot \delta F = \tr (B^T\delta F)$. Then
using the chain rule we get that critical points must verify the
following set of equations:
\begin{eqnarray}\label{derivative}\hspace{-5mm} \ed f^\lambda(F)\cdot\delta
F  & = &  2\, \mathrm{tr}\left[\left(f_1F^T+f_2(I_1F^T-F^TFF^T)
-\frac{\lambda}{2} \det (F) F^{-1}\right)\delta
F\right]\\ \nonumber & = & 0.\\ \nonumber
\det(F) & = & 1\end{eqnarray}

 since
 \begin{eqnarray}\label{traceformula}
 \ed I_1(F) \cdot\delta F & = & 2 \tr (F^T\delta
F)\\
 \label{traceformula2} \ed I_2(F) \cdot\delta F & = & 2 \left[ I_1 \tr \left(F^T\delta
F\right)- \tr \left( F^T F F^T \delta F\right)\right]\\
\label{traceformula3}\ed\det (F)\cdot\delta F & = & \det
(F)\mathrm{tr}(F^{-1}\delta F) .
\end{eqnarray}
(see for instance \cite{Ciarlet} or \cite{MaHu}).

Next proposition gives the expression for the  locked inertia tensor.

\begin{prop} \label{lockedprop}The locked inertia tensor for the $G$-action on $T^*\mathrm{GL}^+(3)$, at a configuration $F\in\mathrm{GL}^+(3)$,
is defined by
\begin{equation}\label{lockedso}
\langle(\widehat{\xi_L}, \widehat{\xi_R}), \II(F) (\widehat{\eta_L},
\widehat{\eta_R})\rangle = T \tr \left[\widehat{\xi_L}^T
\widehat{\eta_L} FF^T+ \widehat{\xi_R}^T F^TF\widehat{\eta_R}-
\widehat{\xi_L}^T F\widehat{\eta_R}F^T - \widehat{\xi_R}^T
F^T\widehat{\eta_L} F
 \right]
\end{equation}
where $T=\frac{4\pi}{15}\rho$  and $\widehat{\xi_i},
\widehat{\eta_i}\in\mathfrak{so}(3)$ for $i=1,2$.

Under the isomorphism \eqref{skewsymm}, the locked inertia tensor is also equivalent to:
\begin{equation}\label{lockedr}
\langle(\xi_L,\xi_R),\II (F) (\eta_L,\eta_R)\rangle=T
\begin{bmatrix}
\xi_L&\xi_R
\end{bmatrix}\begin{bmatrix} \mathbf{i}_S & -2\mathrm{det}\,(F)F^{-T}\\ -2\mathrm{det}\,(F)F^{-1} & \mathbf{i}_C \end{bmatrix} \begin{bmatrix} \eta_L\\
\eta_R  \end{bmatrix}
\end{equation}
where  $S=FF^T$,  $C=F^TF$ and
$\mathbf{i}_A=\mathrm{tr}(A)\mathbf{I}-A$  ($\mathbf{I}$ denotes the
identity matrix).
\end{prop}

\begin{proof} By the  locked inertia tensor definition in Proposition \ref{REthm}, the expression \eqref{infgene}
 for the infinitesimal generators of the $G$-action on $\mathrm{GL}^+(3)$ and the definition \eqref{riem} for the Riemannian metric,  we have
\begin{align*}
\langle(\widehat{\xi_L}, \widehat{\xi_R}),\II(F) (\widehat{\eta_L}, \widehat{\eta_R})\rangle &=  \ll (\widehat{\xi_L}, \widehat{\xi_R})_{\mathrm{GL}^+(3)}(F)\, ,\, (\widehat{\eta_L}, \widehat{\eta_R})_{\mathrm{GL}^+(3)}(F)\gg\\
&=\ll \widehat{\xi_L} F - F \widehat{\xi_R} \, , \, \widehat{\eta_L} F - F \widehat{\eta_R} \gg\\
&= T \tr \left[\left(\widehat{\xi_L}F-F\widehat{\xi_R}\right)^T
\left(\widehat{\eta_L}F-F\widehat{\eta_R}\right)\right].
\end{align*}
Using the fact that $\widehat{\xi_i}$ and $\widehat{\eta_i}$ are
skew-symmetric matrices and the cyclic property of the trace of a
matrix, it is straightforward to obtain expression \eqref{lockedso}.

For the expression \eqref{lockedr} we need some standard properties of the isomorphism \eqref{skewsymm}. In particular,
\begin{eqnarray}
\tr \left(\widehat{\xi}^T\widehat{\eta}\right)= 2\, \xi\cdot\eta&\quad \label{eqisofirst}\\
\tr\left(\widehat{\xi} L\right) = \frac{1}{2} \tr \left(\widehat{\xi} (L-L^T)\right)&\quad  \label{eqisosecond}\\
L\widehat{\xi}+\widehat{\xi}L = \widehat{\mathbf{i}_Lv}&\quad\text{if $L$ is a symmetric matrix} \label{eqisothird}\\
\widehat{L\xi} = \det (L) L^{-T} \widehat{\xi} L^{-1}&\quad\text{if
$L$ is an invertible matrix} \label{eqisofour}
\end{eqnarray}
where $\cdot$ denotes the standard  inner product on $\R^3$. Let us
compute some terms of the expression \eqref{lockedso} since the
other are done similarly
\begin{align*}
\tr \left(\widehat{\xi_L}^T \widehat{\eta_L} F F^T \right)&= \frac{1}{2} \tr\left[\widehat{\xi_L}^T \left(\widehat{\eta_L} FF^T+FF^T\widehat{\eta_L}\right)\right]&\quad\text{(by \eqref{eqisosecond})}\\
&=\frac{1}{2} \tr \left(\widehat{\xi_L}^T \widehat{\mathbf{i}_S\eta_L}\right)&\quad\text{(by \eqref{eqisothird})}\\
&=  \xi_L\cdot \mathbf{i}_S\eta_L&\quad\text{(by \eqref{eqisofirst})}\\
\end{align*}
\begin{align*}
\tr \left(\widehat{\xi_R}^T F^T\widehat{\eta_L} F \right)&=
\frac{1}{\det (F^{-1})} \tr\left[\widehat{\xi_R}^T \widehat{F^{-1} \eta_L}\right]&\quad\text{(by \eqref{eqisofour})}\\
&=2\det (F)  \xi_R\cdot F^{-1} \eta_L &\quad\text{(by \eqref{eqisofirst})}\\
\end{align*}
\end{proof}

As a straightforward consequence we can obtain the momentum of a
relative equilibrium for Dirichlet's model, that is its angular
momentum and circulation.

\begin{cor}\label{momcirc} The momentum for a relative equilibrium with configuration $F$  and velocity-vorticity pair
$(\xi_L,\xi_R)\in\R^3\oplus\R^3$ is
\begin{equation}
\mu=\II(F)(\xi_L,\xi_R) = T\,
\left(\mathbf{i}_S\xi_L-2\det(F)F^{-T}\xi_R\, ,\,
\mathbf{i}_C\xi_R-2\det(F)F^{-1}\xi_L\right).
\end{equation}
That is, the angular momentum and circulation of a Riemann ellipsoid
with configuration by $F$, and angular velocity-vorticity pair
$(\xi_L,\xi_R)$ are given, respectively, by
\begin{eqnarray*}
\mathfrak{j}/T & = & \mathbf{i}_S\xi_L-2\det(F)F^{-T}\xi_R\\
\mathfrak{c}/T & = & \mathbf{i}_C\xi_R-2\det(F)F^{-1}\xi_L
\end{eqnarray*}
\end{cor}

The expression for the twice augmented potential
$V^\lambda_{\xi_L,\xi_R}$ follows now
easily from Proposition \ref{lockedprop}.
\begin{equation}\label{doubleaugmentedV}\begin{array}{ll}V^\lambda_{\xi_L,\xi_R}(F)= & -R\int_0^\infty\frac{ds}{\Delta
(F)}\\ & -T\left(\frac 12\xi_L\cdot\mathbf{i}_S\xi_L+\frac 12\xi_R\cdot
\mathbf{i}_C\xi_R-2\det (F)\xi_L\cdot F^{-T}\xi_R\right)\\
& -\lambda\det (F).\end{array}\end{equation}

\section{Existence conditions for symmetric Riemann ellipsoids}\label{sec existence}

In this section we classify symmetric relative equilibria for
Dirichlet's model. We will treat the spherical case (i.e. a
configuration having three equal principal axes) as a particular
case of a symmetric configuration. From the singular value
decomposition and the definition of the action of
$G=\Z_2\ltimes(\mathrm{SO}(3)\times\mathrm{SO}(3))$ on
$M=\mathrm{GL}^+(3)$ (or on $\mathrm{SL}^+(3)$) it follows that the
stabilizer of a symmetric configuration $F$ is conjugate to the
stabilizer of a diagonal configuration. That is, conjugate to
$\Z_2\ltimes \mathrm{O}(2)_\mathbf{e}^D$ or
$\Z_2\ltimes\mathrm{SO}(3)^D$ if $F$ has 2 or 1 different singular
values, respectively (see \cite{MEres} for a derivation of this
result).
Actually, if the configurations are diagonal, these are exactly
their stabilizers. Here, $K^D$ denotes the diagonal embedding of
$K\subset \mathrm{SO}(3)$ in $\mathrm{SO}(3)\times\mathrm{SO}(3)$
and $\mathrm{O}(2)_\mathbf{e}$ is the subgroup of $\mathrm{SO}(3)$
generated by all the  rotations $R_\theta\in
\mathrm{SO}(2)_\mathbf{e}$ around a given axis $\mathbf{e}$ in
$\R^3$ and a rotation, $\Pi_\mathbf{e^\perp}$, by $\pi$ around an
axis $\mathbf{e^\perp}$ perpendicular to $\mathbf{e}$. In case of
the diagonal configuration $F=\mathrm{diag}(a,a,c)$, $R_\theta$ is
the rotation matrix by and angle $\theta$ around $(0,0,1)$  and
$\Pi_\mathbf{e^\perp}$ can be chosen to be $\mathrm{diag}\,
(1,-1,-1)$. We  introduce the following subgroups:
\begin{itemize}

\item $\widetilde{\mathrm{SO}(2)_{\mathbf{e}}\times\mathrm{SO}(2)_{\mathbf{e}}}$, generated by elements $(e;R_{\theta_1},R_{\theta_2})$, with $R_{\theta_{1,2}}\in\mathrm{SO}(2)_{\mathbf{e}}$ and $(\sigma; \Pi_\mathbf{e^\perp},\Pi_\mathbf{e^\perp})$,

\item $\widetilde{\mathrm{O}(2)_\mathbf{e}}$, generated by elements  $(e;R_{\theta},R_{\theta})$, with $R_\theta\in\mathrm{SO}(2)_{\mathbf{e}}$ and $(\sigma; \Pi_\mathbf{e^\perp},\Pi_\mathbf{e^\perp})$,

\item $\mathbb{Z}_2(\mathbf{e})$, the cyclic group isomorphic to $\mathbb{Z}_2$ generated by the element $(e;\Pi_\mathbf{e},\Pi_\mathbf{e})$.

\item More generally, if $K$ is a subgroup of $\mathrm{SO}(3)\times\mathrm{SO}(3)$, we denote also by $K$ the subgroup of $\Z_2\ltimes(\mathrm{SO}(3)\times\mathrm{SO}(3))$ generated by elements $(e;k)$, with $k\in K$.

\end{itemize}

Note that since we are going to impose the constraint $F\in\SL$,  we
will consider only two kinds of symmetric configurations,
especifically:
\begin{itemize}
\item spherical: $F=\diag (1,1,1)$,
\item spheroidal: $F=\diag (a,a,c)$, with $a^2c=1$.
\end{itemize}

To find all the possible Riemann ellipsoids with symmetric
configurations, we will have to solve \eqref{existence2augmented}
with $F$ of the above forms and different pairs $(\xi_L,\xi_R)$. The
possible solutions are summarized in the following theorem.

\begin{thm} \label{theorem existence}
The relative equilibria for Dirichlet's model of a self-gravitating
fluid mass are:

\begin{itemize}
\item[(i)] The {\bf spherical equilibrium} with spherical configuration $F=\mathrm{diag}(1,1,1)$, velocity-vorticity pair $(0,0)$ and
Lagrange multiplier $\lambda=2V_1+4V_2$. Its corresponding momentum and isotropy groups are
$$\mu=(\mathbf{j},\mathbf{c})=(0,0),\quad G_\mu=\Z_2\ltimes(\mathrm{SO}(3)\times \mathrm{SO}(3)),\quad G_F=G_{p_F}=\mathbb{Z}_2\ltimes\mathrm{SO}(3)^D.$$
\item[(ii)] The family of {\bf MacLaurin spheroids} which have  oblate
spheroidal con\-fi\-gu\-ra\-tions $F=\mathrm{diag}(a,a,c)$ (with
$c<a$) and angular velocity and vorticity parallel to the axis of
symmetry $\mathbf{e_3}$. In terms of the parameter $\Omega$ defined
by $\Omega\,\mathbf{e_3}=\xi_L- \xi_R$,  this family is
characterized by $\lambda=2(1-e^2)^{2/3}V_1+4(1-e^2)^{1/3}V_2$ and
the following constraint between $\Omega$ and the eccentricity $e$:
\begin{equation}\label{maclaurin formula}
\frac{\Omega^2}{\pi \rho G} = 2\frac{\sqrt{1-e^2}}{e^3} (3-2
e^2)\arcsin e - \frac{6}{e^2} (1-e^2).
\end{equation}
Its corresponding momentum and isotropy groups are:
$$\mu  =  (\mathbf{j},\mathbf{c})=2(1-e^2)^{-1/3}T\Omega(\mathbf{e_3},-\mathbf{e_3}),$$
\begin{align*}
 G_F=\Z_2\ltimes\mathrm{O}(2)_{\mathbf{e_3}}^D,\quad&
G_\mu
=\widetilde{\mathrm{SO}(2)_{\mathbf{e_3}}\times\mathrm{SO}(2)_{\mathbf{e_3}}},\quad&
G_{p_F}=\widetilde{\mathrm{O}(2)_\mathbf{e_3}}.\end{align*}

\item[(iii)]
Two branches of {\bf transversal spheroids} which have prolate
spheroidal configurations $F=\mathrm{diag}(a,a,c)$ (with $c>a$). We
distinguish the two branches of this family with the signs $+$ and
$-$. These branches are characterized by the Lagrange multiplier
$\lambda = 2 ((1-e^2)^{1/3} V_1+ (1-e^2)^{-1/3}(e^2-2)V_2)$, the
velocity-vorticity pair $(\xi_L,\xi_R)_\pm=\w_\pm (\mathbf{n} ,
f_{\pm}\mathbf{n})$ with $f_{\pm} = \frac{1\pm e}{\sqrt{1-e^2}}$
(where $\mathbf{n}$ is a unit vector perpendicular to $\etres$) and
the following constraints between $\w_\pm$ and the eccentricity $e$:
\begin{equation}\label{transversal formula}\frac{\w^2_\pm}{\pi\rho G}   =  \mp \frac{(e\mp 1)^2 (e\pm 1)}{e^5}\left(3 e + (e^2-3)\mathrm{arctanh}\, e\right).\end{equation}
The corresponding momentum and isotropy groups are:
$$ \mu_{\pm}  =  T\w_{\pm}\left(-\frac{e(e\pm 2)}{(1-e^2)^{2/3}}\mathbf{n},\pm \frac{e(e\mp 2)}{(e\mp 1)(1-e^2)^{1/6}}\mathbf{n}\right)$$
\begin{align*}
G_F=\Z_2\ltimes\mathrm{O}(2)_{\mathbf{e_3}}^D,\quad & G_\mu
=\mathrm{SO}(2)_{\mathbf{n}}\times\mathrm{SO}(2)_{\mathbf{n}},\quad
& G_{p_F}=\mathbb{Z}_2(\mathbf{n}).\end{align*}

\end{itemize}

\end{thm}
Before proving the theorem we remark that formula \eqref{maclaurin
formula}  has already been obtained by MacLaurin in 1742, as it is claimed in
page 4 of Chandrasekhar's book \cite{Chand}.
\begin{proof}
First, using \eqref{doubleaugmentedV}, \eqref{first derivatives} and
\eqref{derivative}, it is easy to see that the general conditions
\eqref{existence2augmented} are equivalent to
\begin{eqnarray}
\label{REcond1} 0 & = & 2V_1F^T+2V_2(I_1F^T-F^TFF^T)-\lambda\det(F)F^{-1} \\ \nonumber & & -T\,\left[(\|\xi_L\|^2+\|\xi_R\|^2)F^T-F^T(\xi_L\otimes\xi_L)-(\xi_R\otimes\xi_R)F^T\right.\\
\nonumber & &\left.+2\det(F)\left((F^{-1}\xi_L\otimes F^{-T}\xi_R)
-(\xi_L\cdot F^{-T}\xi_R)F^{-1}\right)\right]\\
\label{REcond2} 1 & = & \det(F)
\end{eqnarray}

\paragraph{spherical case:} If $F=\mathbf{I}$, then \eqref{REcond1}, \eqref{REcond2} are simply
$$\left(2V_1+4V_2-\lambda-T\left[\|\xi_L\|^2+\|\xi_R\|^2-2\xi_L\cdot\xi_R\right]\right)
\mathbf{I}+T\left[\xi_L\otimes\xi_L+\xi_R\otimes\xi_R-2\xi_L\otimes\xi_R\right]=0$$
Taking $\xi_L=(\xi_{L,1},\xi_{L,2},\xi_{L,3})$ and the same sorte of
notation for $\xi_R$, the off-diagonal terms of this expression are
independent of $V_1,V_2$ and $\lambda$, and  equivalent to the
following 6 equations:
\begin{align*}
(\xi_{L,1}-\xi_{R,1})(\xi_{L,2}-\xi_{R,2})=0, & \quad & \xi_{R,1}\xi_{L,2}=\xi_{R,2}\xi_{L,1}\\
(\xi_{L,1}-\xi_{R,1})(\xi_{L,3}-\xi_{R,3})=0, & \quad & \xi_{R,1}\xi_{L,3}=\xi_{R,3}\xi_{L,1}\\
(\xi_{L,2}-\xi_{R,2})(\xi_{L,3}-\xi_{R,3})=0, & \quad & \xi_{R,2}\xi_{L,3}=\xi_{R,3}\xi_{L,2}\\
\end{align*}
It follows then that $\xi_L=\xi_R$. Recall from Theorem \ref{REthm}
that the momentum of a relative equilibrium with configuration $x$
and velocity $\xi$ is given by $\mu=\II (x)(\xi)$. Then, from
\eqref{lockedr} we have $\mu=(\mathbf{j},\mathbf{c})=(0,0)$.
Therefore, $G_F=\mathbb{Z}_2\ltimes\mathrm{SO}(3)^D$, $G_\mu=G$,
$G_{p_F}=G_\mu\cap G_F=G_F$.

Now, noting that $(\xi_L,\xi_R)\in \g_{p_F}$ if $\xi_L=\xi_R$, and
that the velocity of a relative equilibrium is defined only up to
addition of elements in $\g_{p_F}$, the relative equilibrium
$(\mathbf{I},(\xi_L,\xi_R))$ is the same as $(\mathbf{I},(0,0))$.
Then the relative equilibrium conditions are satisfied with
$\lambda=2V_1+4V_2$.

\paragraph{spheroidal case:} We now consider $F=\diag (a,a,c)$ with $a^2c=1$. Since in the $\langle \mathbf{e_1},\mathbf{e_2}\rangle$ plane all directions are equivalent, we can assume without loss of generality that $\xi_{L,1}=0$. Now, conditions \eqref{REcond1}, \eqref{REcond2} are equivalent to the system
\begin{align}
\label{sa1} 0  &=  2aV_1+2a^3V_2+2ac^2V_2-ac\lambda-aT(\xi_{R,2}^2+\xi_{R,3}^2+\xi_{L,2}^2+\xi_{L,3}^2)\\
\nonumber  &\qquad +2T(c\xi_{R,2}\xi_{L,2}+a\xi_{R,3}\xi_{L,3})\\
\label{sa2} 0 & =  2aV_1+2a^3V_2+2ac^2V_2-ac\lambda -aT(\xi_{R,1}^2+\xi_{R,3}^2+\xi_{L,3}^2)\\
\nonumber  &\qquad+2aT\xi_{R,3}\xi_{L,3}\\
\label{sa3} 0 & =  2cV_1+4a^2cV_2-a^2\lambda -cT(\xi_{R,1}^2+\xi_{R,2}^2+\xi_{L,2}^2)+2aT\xi_{R,2}\xi_{L,2}\\
\label{sb1} 0 &=\xi_{R,1} \xi_{R,2} =\xi_{R,1}\xi_{R,3}=\xi_{R,1}\xi_{L,2}=\xi_{R,1}\xi_{L,3}\\
\label{sb2} 0 & =  cT\xi_{R,2}\xi_{R,3}-2aT\xi_{R,3}\xi_{L,2}+aT\xi_{L,2}\xi_{L,3}\\
\label{sb3}  0 & =  aT\xi_{R,2}\xi_{R,3}-2aT\xi_{R,2}\xi_{L,3}+cT\xi_{L,2}\xi_{L,3}\\
\label{sc} 1 & =  a^2c
\end{align}
 Note that these
equations imply that $\xi_{R,1}=0$. Indeed, if $\xi_{R,1}\neq 0$
then \eqref{sb1} implies $\xi_{R,2}=\xi_{R,3}=\xi_{L,2}=\xi_{L,3}=0$
and  so \eqref{sa1}and \eqref{sa2} imply $\xi_{R,1}=0$ which is a
contradiction.  We will now proceed systematically considering four
main cases: (i) $\xi_L=\xi_R=0$, (ii) $\xi_{L,2}=0$, (iii)
$\xi_{L,3}= 0$ and $\xi_{L,2}\neq0$, and (iv) $\xi_{L,3}\neq 0$ and
$\xi_{L,2}\neq0$.

\noindent (i) If $\xi_L=\xi_R=0$, then \eqref{sa2}, \eqref{sa3} and
\eqref{sc} are the only non-trivial conditions, which are equivalent
to
$$a^2c=1,\quad \lambda=\frac{c}{a^2}(2V_1+4a^2V_2),\quad (a^6-1)(V_1+a^2V_2)=0.$$
As $V_1$ and $V_2$ are positive and $0<a\neq 1$ it follows from the last
equation that there is no solution.

\noindent (ii) If  $\xi_{L,2}=0$, then from \eqref{sa1} and
\eqref{sa2} we have $\xi_{R,2}=0$ and so \eqref{sa3} and \eqref{sc}
give $\lambda=2c^2V_1+4cV_2$.

Let $\Omega=\xi_{L,3}-\xi_{R,3}$. Then  the remaining non-trivial
equations, \eqref{sa1}, \eqref{sa2} and \eqref{sc},  give
\begin{equation}\label{omegaS}\Omega^2=\frac 2T\frac{a^2-c^2}{a^2}(V_1+a^2V_2).\end{equation}
As $V_1$ and $V_2$ are positive, then last equality implies that the
spheroidal configuration $F$ is oblate, that is $a>c$. The
eccentricity of the spheroid is $e^2= 1- \frac{c^2}{a^2}$ and  $a^2
= (1-e^2)^{-1/3}$.  Using the relations \eqref{first derivatives}
for the partial derivatives of the self-gravitating potential, $V_1$
and $V_2$,  the expression $J_O$ for the integrals $J(k,r)$ in the
oblate case given by \eqref{JO}  and  $\frac{R}{T} = 2 \pi \rho G$,
then  the equation \eqref{omegaS} is equivalent to
\begin{align*}
\Omega^2 &= \frac{R}{T} e^2  \left(J_O(3,2)+ (1-e^2)^{-1/3} J_O(3,1)\right)\\
&= 2 \pi\rho G e^2  \left(J_O(3,2)+ (1-e^2)^{-1/3} J_O(3,1)\right).
\end{align*}
 One can easily compute   the definite integrals $J_O(3,2) $ and $J_O(3,1)$, although we  avoid to display their expressions since they are quite lengthy.  However   the expression $J_O(3,2)+ (1-e^2)^{-1/3} J_O(3,1)$ is
 $$J_O(3,2)+ (1-e^2)^{-1/3} J_O(3,1) = 3\, \frac{e^2-1}{e^4}+ \frac{3-2e^2}{e^5}\sqrt{1-e^2}\arcsin e,
 $$
from which \eqref{maclaurin formula} follows. From Corollary
\ref{momcirc} it is trivial to obtain that the momentum of this
relative equilibrium. Hence, using \eqref{momentumcarac}, the
appropriate isotropy groups are also straightforward.

\noindent (iii) If $\xi_{L,3}= 0$ and $\xi_{L,2}\neq 0$, it follows
from \eqref{sb2} and \eqref{sb3} that $\xi_{R,3}=0$. Then
\eqref{sa2} and \eqref{sc} give
$$a^2c=1,\quad \lambda=2a^2(V_1+ (a^2+c^2)V_2).$$
We can  set $\mathbf{n}=\mathbf{e_2}$ and
$(\xi_L,\xi_R)=\w(\mathbf{n},f\mathbf{n})$. So,  substituting the
above value of $\lambda$ into \eqref{sa1} and \eqref{sa3} these
equations are
\begin{align}
\label{tr1} 0 & =  (a-2cf+af^2)T\w^2\\
\label{tr2} 0 & =
2(c^2-a^2)V_1+2a^2(c^2-a^2)V_2-(c^2f^2-2acf+c^2)T\w^2
\end{align}
From \eqref{tr1} we obtain the solutions
$f_{\pm}=\frac{c\pm\sqrt{c^2-a^2}}{a}$, from which follows that the
spheroids must be prolate $(c>a)$. In terms of the eccentricity
$e^2=1-\frac{a^2}{c^2}$ we have
\begin{equation*}f_{\pm}=\frac{1\pm e}{\sqrt{1-e^2}}.\end{equation*}
Therefore, the equation \eqref{tr2} gives
$$\w_{\pm}^2=\frac{2(c^2-a^2)(V_1+a^2V_2)}{T(c^2f_{\pm}^2-2acf_{\pm}+c^2)}= \frac{1\mp e}{T} \left(V_1+(1-e^2)^{1/3} V_2)\right).$$
Substituting in this expression $V_1=\frac{R}{2}J_P(3,2)$ and
$V_2=\frac{R}{2}J_P(3,1)$, as well as $R= 2 \pi \rho G T$ gives
\eqref{transversal formula}. As before, using the expression of the
locked inertia tensor and \eqref{momentumcarac} the remaining results follow.

\noindent (iv) In this case we have $\xi_{L,2}\neq 0$,
$\xi_{L,3}\neq 0$  and $\xi_{L,1}= \xi_{R,1}=0$. Note that from
\eqref{sb2} and \eqref{sb3} one should also have $\xi_{R,2}\neq 0$
and $\xi_{R,3}\neq 0$. So,  let $\xi_{L,2} = g \xi_{R,2} $ and
$\xi_{L,3} = h \xi_{R,3}$ for some reals $g,h\neq 0$. Then, using
also \eqref{sc}, the equations \eqref{sb2} and \eqref{sb3}  reduce
to
\begin{align*}
(1-2 a^3 g + a^3  gh)\xi_{R,2}\xi_{R,3}&=0\\
(a^3-2 a^3 h +  gh)\xi_{R,2}\xi_{R,3}&=0.
\end{align*}

 These equations have solutions
\begin{equation}\label{no solution}(h_\pm,g_\pm)=\left(\frac{5a^6-1\pm\sqrt{1-10 a^6+9a^{12}}}{4a^6} , \frac{1+ 3a^6\pm\sqrt{1-10 a^6+9a^{12}}}{4a^3}\right).
\end{equation}

 Comparing \eqref{sa1} and  \eqref{sa2} using $\xi_{L,2} = g \xi_{R,2} $,
$\xi_{L,3} = h \xi_{R,3}$ and \eqref{sc} we obtain $-a(1+g^2)+2cg=0$
and so $g_\pm$ must be $\frac{1+\sqrt{1-a^6}}{a^3}$ or
$\frac{1-\sqrt{1-a^6}}{a^3}$, but this is a contradiction with
\eqref{no solution}. So there is no solution for the above system.
 \end{proof}

\begin{rem} \label{transversal adjoint}

Note that the functions $f_+$ and $f_-$ appearing in the
characterization of transversal spheroids are inverse of each other.
Therefore the two families of transversal spheroids belong to a
single orbit of the symmetry group. Indeed, the $\mathbb{Z}_2$
symmetry interchanges the $+$ and $-$ families, since $\sigma\cdot
(F;\w_+\mathbf{n},\w_+
f_+\mathbf{n})=(F^T;\w_+f_+\mathbf{n},\w_+\mathbf{n})=(F;\w_-\mathbf{n},\w_-f_-\mathbf{n})$,
as it follows from their definitions that $\w^2_+/\w^2_-=f_-/f_+$.
\end{rem}

\begin{rem} \

\noindent 1. Theorem~\ref{theorem existence} is in agreement with
Riemann's classification of ellipsoidal fi\-gu\-res of equilibrium
for Dirichlet's model of self-gravitating fluid masses. That is,
these ellipsoidal figures of equilibrium must lie in one of the
following categories: (a) the case of a uniform rotation with no
internal motion (or uniform vorticity and no rotation); (b) the case
when the directions of the angular velocity $\xi_L$ and vorticity
$\xi_R$ are the same and coincide with a principal axis of the
ellipsoid (also known as ellipsoids of type S); (c) The case when
the angular velocity and vorticity are not parallel but lie in the
same principal plane.

In particular, we show that for Dirichlet's model it is not possible
to obtain relative equilibria with spheroidal configurations
belonging to category (c).

\noindent 2.  The existence of transversal spheroids is referred in
Chandrasekhar's book \cite{Chand} (see for instance page 143),
however their study is not present in the classical works of
Liapunov \cite{Liapunov} and Poincar\'e \cite{Poincare}.
\end{rem}

 \section{Stability conditions for symmetric Riemann ellipsoids}\label{sec stability}

In this section we apply the singular version  of the reduced
energy-momentum method introduced in  \cite{miguelstab} in order to
deduce the stability of the symmetric relative equilibria obtained
in  Theorem~\ref{theorem existence}. In order to apply this method
it is essential to compute the second derivative of the twice
augmented potential $V_{\xi_L,\xi_R}^\lambda$. Next lemma gives that
result.

 \begin{lemma}\label{second derivative augmented} If $F$ is a critical point of the twice augmented potential
 $$V_{(\xi_L,\xi_R)}^\lambda = V(I_1(F),I_2(F))-\frac{1}{2} \begin{bmatrix}\xi_L& \xi_R\end{bmatrix} \II(F) \begin{bmatrix}\xi_L\\ \xi_R\end{bmatrix} -\lambda \det(F),$$
for $(\xi_L,\xi_R)\in\R^3\times\R^3$ and $\II(F)$ as in Proposition~\ref{lockedprop} and $A,B\in T_F\mathrm{GL}^+(3)$, then
\begin{align*}
\ed_F^2V_{(\xi_L,\xi_R)}^\lambda(A,B)&=\ed_F^2V(A,B)-\frac 12
\begin{bmatrix}\xi_L& \xi_R\end{bmatrix}
(\mathbf{D}_F^2\II(A,B))\begin{bmatrix}\xi_L\\ \xi_R\end{bmatrix}\\
& -\lambda\det(F)\left(\tr(F^{-1}B)\tr(F^{-1}A)-\tr(F^{-1}BF^{-1}A)\right)
\end{align*}
where
\begin{align*}
\ed_F^2V(A,B) & = 2   \tr (B^T A) \left(V_1+ I_1 V_2\right)\\
&- 2 \tr \left(BF^TFA^T+ FB^TFA^T+FF^TBA^T\right) V_2\\
&+ 4 \tr( F^TA)  \tr( FB^T)\left(V_2+V_{11}+2 I_1 V_{12}+ I_1^2 V_{22}\right)\\
&- 4 \tr( FF^TFB^T) \tr( F^TA) \left(V_{12}+ I_1 V_{22}\right)\\
&- 4 \tr( F^TFF^TA) \tr( FB^T) \left(V_{12}+ I_1 V_{22}\right)\\
&+ 4 \tr( F^TFF^TA) \tr( FF^TFB^T)V_{22}
\end{align*}
and
\begin{align*}
\begin{bmatrix}\xi_L& \xi_R\end{bmatrix}
\mathbf{D}^2_F\II(A,B)\begin{bmatrix}\xi_L\\ \xi_R\end{bmatrix} & =
T  \tr
(4\widehat{\xi_R}B^T\widehat{\xi_L}A-2\widehat{\xi_L}^2AB^T-2\widehat{\xi_R}^2B^TA)
\end{align*}
 \end{lemma}
 \begin{proof} We will just sketch the computation of $\ed_F^2V(A,B)$.

Recall from the proof of Theorem \ref{theorem existence}, that
  \begin{align*}
 \ed V(F)\cdot A  = 2  (V_1+ I_1 V_2)\tr (F^TA)- 2 V_2\tr (F^TFF^TA).
 \end{align*}
Differentiating again using the expressions for $ \ed I_1(F)\cdot A$
and $\ed I_2(F)\cdot A$ given in \eqref{traceformula} and
\eqref{traceformula2} and the chain rule the result follows.

For the expression $\mathbf{D}^2_F\II(A,B)$ we differentiate the
expression \eqref{lockedso}, which in this case takes the form
 \begin{align*}
 \left\langle(\widehat{\xi_L}, \widehat{\xi_R}), \II(F) (\widehat{\xi_L}, \widehat{\xi_R})\right\rangle & =  T \tr \left[ 2 \widehat{\xi_R} F^T\widehat{\xi_L} F - F^TF\widehat{\xi_R}^2-FF^T\widehat{\xi_L}^2\right]
 \end{align*}

  Then, applying standard properties of the trace, we get
   $$
 \left\langle(\widehat{\xi_L}, \widehat{\xi_R}), (\mathbf{D}\II(F)\cdot A) (\widehat{\xi_L}, \widehat{\xi_R})\right\rangle =  T \tr \left[ 4 \widehat{\xi_R} A^T\widehat{\xi_L} F - 2 A^TF\widehat{\xi_R}^2-2AF^T\widehat{\xi_L}^2\right].
$$
The expression for $\mathbf{D}^2_F\II(A,B)$ stated follows now easily. Finally, using
 \eqref{traceformula3} to differenciate the expression
\begin{align*}
\ed \det(F)\cdot A&= \det (F)\tr(F^{-1}A)
\end{align*}
we get
\begin{align*}
\ed^2_F\left(\det(F)\right)(A,B)&= \det (F)\left[\tr(F^{-1}B)\tr(F^{-1}A)-\tr(F^{-1}BF^{-1}A)\right].
\end{align*}

  \end{proof}

            \subsection{Spherical equilibrium}

We now study the stability of the spherical equilibrium. Notice from
the outline of the method in Section 3 that for this equilibrium whe
have that $\q^\mu$, the correction term and the Arnold form are all
trivial, as well as the velocity-vorticity pair $(\xi_L,\xi_R)$. As
a consequence, $\Sigma^{\SL}_{\mathrm{int}}=\mathbf{S}^{\SL}$, the
orthogonal complement to the $G$-orbit at the identity in $\SL$.
Hence, to conclude stability of the spherical equilibrium we need to
study the definiteness of
$$\ed^2_\mathbf{I}V^\lambda_{(0,0)}\restr{\Sl^{\SL}}.$$

\begin{thm}\label{spherical stability} For Dirichlet's model, the spherical equilibrium is nonlinearly $G$-stable.
\end{thm}
\begin{proof}
Recall that $T_\mathbf{I}\SL$ is the space of traceless matrices. Also, the infinitesimal action of $\g$ on $\mathrm{GL}^+(3)$ at $\mathbf{I}$ is $(\xi_L,\xi_R)_{\mathrm{GL}^+(3)}(\mathbf{I})= \widehat{\xi_L}-\widehat{\xi_R}$. Then
$$\Sl^{\SL}=\{A\in T_\mathbf{I}\SL\,:\,\tr (A\widehat{\xi_L}-A\widehat{\xi_R})=0\quad\forall\,\xi_L,\xi_R\in\R^3\}.$$
Therefore $\Sl^{\SL}$ is the space of traceless symmetric matrices. That is, matrices of the form
$$A=\begin{bmatrix}
a_{11} & a_{12} & a_{13}\\
a_{12} & a_{22} & a_{23}\\
a_{13} & a_{23} & -(a_{11}+a_{22})
\end{bmatrix}.$$
We fix a basis  for  $\Sl^{\SL}$ with respect to which the components of $A$ are $(a_{11},a_{22},a_{12},a_{13},a_{23})$.

By Lemma~\ref{second derivative augmented},  the expression of $\ed^2_\mathbf{I}V^\lambda_{(0,0)}\restr{\Sl^{\SL}}$, with $\lambda = 2V_1+4V_2$ as given in Theorem \ref{theorem existence}, reduces to:
\begin{align*}
\ed^2_\mathbf{I}V^\lambda_{(0,0)}\restr{\Sl^{\SL}}(A,B)= 4 (V_1+V_2) \tr(BA),
\end{align*}
where we have applied the fact that  $A$ and $B$ are traceless
symmetric matrices.
Therefore in this basis we have
$$\ed^2_\mathbf{I}V^\lambda_{(0,0)}\restr{\Sl^{\SL}}=4(V_1+V_2)\begin{bmatrix}
2 & 1 & 0 & 0 & 0 \\
1 & 2 & 0 & 0 & 0 \\
0 & 0 & 2 & 0 & 0 \\
0 & 0 & 0 & 2 & 0 \\
0 & 0 & 0 & 0 & 2
\end{bmatrix} $$
As   $V_1+V_2 $ is positive, the
eigenvalues of this matrix are $12(V_1+V_2), 4(V_1+V_2)$ and $8(V_1+V_2)$ with multiplicities $1,1$ and $3$ respectively. These are all positive, therefore  the spherical equilibrium is $G$-stable.
\end{proof}

            \subsection{MacLaurin spheroids}

We now study the nonlinear stability of MacLaurin spheroids in the setup of previous sections (Theorem \ref{theorem existence}). As it has been stated, a MacLaurin spheroid has an oblate configuration which, with no loss of generality, we suppose diagonal. This  configuration
is uniquely characterized by the eccentricity $e\in(0,1)$. In order to apply Theorem~\ref{energy miguel}
one needs first to split $\g=\R^3\oplus\R^3$ according to \eqref{split}, that is as
$$\g_\mu =\g_{p_F}\oplus \mathfrak{p}\qquad\text{and}\qquad \g = \g_F\oplus\mathfrak{p}\oplus \mathfrak{t}.
$$
Recall that for the MacLaurin spheroid one has $G_F=\Z_2\ltimes\mathrm{O}(2)^D$ and $G_\mu = \widetilde{\mathrm{SO}(2)_\mathbf{e_3}\times \mathrm{SO}(2)_\mathbf{e_3}}$.  One can then choose the following ordered orthonormal bases (with respect to the Euclidean product in $\R^3\oplus\R^3$)
for each of the spaces of the above splitting: If we define $h=\frac{1}{\sqrt{2}}(\mathbf{e_3},\mathbf{e_3}),\,p=\frac{1}{\sqrt{2}}(\mathbf{e_3},-\mathbf{e_3}), t_1=(\mathbf{e_1}, 0),\,t_2=(0, \mathbf{e_1}),\,t_3=(\mathbf{e_2}, 0),\,t_4= (0, \mathbf{e_2})$, then
\begin{align*}
\g_F&=\operatorname{span}\{ h\}\\
\mathfrak{p}&=\operatorname{span}\{ p\}\\
\mathfrak{t}&=\operatorname{span}\{ t_1,t_2,t_3,t_4\}.
\end{align*}
It is straightforward to check that these subspaces are invariant for $G_{P_F}=\widetilde{\mathrm{O}(2)_\mathbf{e_3}}$.
The orthogonal velocity $\xi^\perp$ for the MacLaurin relative equilibrium is the orthogonal projection of the velocity $\xi$ onto $\mathfrak{p}$.  Then,
\begin{align*}
(\xi_L,\xi_R)^\perp &=\frac{1}{2} (\xi_{L,3}\etres, \xi_{R,3}\etres)\cdot (\etres, -\etres) (\etres, -\etres)\\
&= \frac{1}{2} (\xi_{L,3}-\xi_{R,3}) (\etres, -\etres)=\frac{\Omega}{2}(\etres, -\etres)=\frac{\Omega}{\sqrt{2}} p
\end{align*}
where $\Omega$ must satisfy \eqref{maclaurin formula}. As already
defined, $\widehat{\II}_0$ is the restriction of $\II$ to
$\mathfrak{p}\oplus\mathfrak{t}$. The locked inertia matrix for the
configuration $F=\diag (a,a,c)$ is, according to \eqref{lockedr},
$$\II(F) =T \begin{bmatrix}
D_1&D_2\\
D_2&D_1
\end{bmatrix}
$$
where $D_1= \diag (a^2+c^2, a^2+c^2, 2 a^2)$ and $D_2 = - \diag
(2ac, 2ac, 2a^2)$. It is now straighforward to obtain the
$\widehat{\II}_0$ matrix with respect to the basis
$(p,t_1,t_2,t_3,t_4)$. That is
$$\widehat{\II}_0 = T \begin{bmatrix}
4a^2&0&0&0&0\\
0&a^2+c^2&-2ac&0&0\\
0&-2ac&a^2+c^2&0&0\\
0&0&0&a^2+c^2&-2ac\\
0&0&0&-2ac&a^2+c^2\\
\end{bmatrix}
$$
Or, in terms of the eccentricity
\begin{equation}\label{Pinot}
\widehat{\II}_0= T\begin{bmatrix}
\frac{4}{\sqrt[3]{1-e^2}} & 0 & 0 & 0 & 0\\
0 & \frac{2-e^2}{\sqrt[3]{1-e^2}} & -2\sqrt[6]{1-e^2} &
0 & 0\\
0 & -2\sqrt[6]{1-e^2}  & \frac{2-e^2}{\sqrt[3]{1-e^2}} &
0 & 0 \\
0 & 0 & 0 &\frac{2-e^2}{\sqrt[3]{1-e^2}}  & -2\sqrt[6]{1-e^2} \\
0 & 0 & 0 & -2\sqrt[6]{1-e^2}  & \frac{2-e^2}{\sqrt[3]{1-e^2}}
\end{bmatrix}
\end{equation}

We can use $\rII$ and $(\xi_L,\xi_R)^\perp$ to compute the momentum of a MacLaurin spheroid. Indeed $$\mu = \rII (\xi_L,\xi_R)^{\perp} =  \frac{\Omega}{\sqrt{2}} \widehat{\II}_0 (p) = \frac{2 \sqrt{2}T\Omega }{(1-e^2)^{1/3}} p $$
which is of course the same as the value obtained in Theorem \ref{theorem existence} under the identification $\g\simeq\g^*$ induced by the Euclidean product in $\R^3\oplus\R^3$.

In order to apply Theorem~\ref{energy miguel} we need to verify that
the singular Arnold form is nondegenerate.

\begin{prop} For a MacLaurin spheroid $\q^\mu=\mathfrak{t}$ and the Arnold form, defined in \eqref{Arnold form}, is positive definite for all eccentricities.
\end{prop}
\begin{proof} Recall that the Arnold form $\Ar: \mathfrak{q}^\mu\times \mathfrak{q}^\mu\rightarrow \R$ is defined by:
$$
\Ar (\gamma_1, \gamma_2) = \langle\ad^*_{\gamma_1}\mu\, ,\, \Lambda (F,\mu) (\gamma_2) \rangle,
$$
where
\begin{equation*}
\Lambda(F,\mu)(\gamma)=\rII^{-1}\left(\ad^*_{\gamma}\mu\right)+\Proj_{\mathfrak{p}^*\oplus\mathfrak{t}^*}\left[\ad_{\gamma}\left(\rII^{-1}\mu\right)\right]
\end{equation*}

First, we compute the space $\q^\mu$. Notice the following relations
for the adjoint representation of $G$:
\begin{equation}\label{severaladjoint}\ad_{t_1}p= \frac{-1 }{\sqrt{2}}t_3,\quad \ad_{t_2}p= \frac{1 }{\sqrt{2}}t_4,\quad
\ad_{t_3}p= \frac{1 }{\sqrt{2}}t_1, \quad \ad_{t_4}p= \frac{-1
}{\sqrt{2}}t_2,\end{equation} and
\begin{equation}\label{severaladjoint2}\begin{array}{ll}
\ad_{t_1}t_2 & = \ad_{t_1}t_4  =\ad_{t_2}t_3 =\ad_{t_3}t_4= 0,\\
\ad_{t_1}t_3 & = \frac{1}{\sqrt{2}}(h+p),\\
\ad_{t_2}t_4 & = \frac{1}{\sqrt{2}}(h-p).
\end{array}
\end{equation}
Also, recall that under our identification $\g\simeq\g^*$ we have
$\ad^*_\gamma\rho=-\ad_\gamma\rho$, for $\gamma,\in\g,\rho\in\g^*$,
and where $\rho$ in the right hand side is identified with an element of $\g$.
Therefore $\Proj_{\g_F}\left(\ad^*_{t_i}\mu\right)=0$ for
$i=1,2,3,4$, hence $\q^\mu=\mathfrak{t}$.
 As $  \frac{\Omega}{\sqrt{2}}
\widehat{\II}_0 (p)= \mu$ then $\rII^{-1}\left(\mu\right) =
\frac{\Omega}{\sqrt{2}} p$.  Then, from \eqref{severaladjoint} we obtain
\begin{align*}
\ad_{t_1}\left(\rII^{-1}\mu\right)&= - \frac{\Omega}{2}  t_3&
\ad_{t_2}\left(\rII^{-1}\mu\right)&=  \frac{\Omega}{2}  t_4\\
\ad_{t_3}\left(\rII^{-1}\mu\right)&=  \frac{\Omega}{2}  t_1&
\ad_{t_4}\left(\rII^{-1}\mu\right)&=  -\frac{\Omega}{2}  t_2
\end{align*}

The inverse of the matrix \eqref{Pinot} is not difficult to compute. Here we just state the values of $\rII^{-1}\left(\ad^*_{w}\mu\right) = -\rII^{-1}\left(\ad_{w}\mu\right)$ on vectors $w$ of the fixed basis:
\begin{align*}
\rII^{-1}\left(\ad^*_{t_1}\mu\right)& = \frac{2T\Omega}{(1-e^2)^{1/3}}\rII^{-1} (t_3)=\frac{-2 (e^2-2)\Omega}{e^4}t_3+\frac{4\sqrt{1-e^2}\Omega}{e^4}t_4\\
\rII^{-1}\left(\ad^*_{t_2}\mu\right)&=-\frac{2T\Omega}{(1-e^2)^{1/3}}\rII^{-1} (t_4)= \frac{-4\sqrt{1-e^2}\Omega}{e^4}t_3+\frac{2 (e^2-2)\Omega}{e^4}t_4\\
\rII^{-1}\left(\ad^*_{t_3}\mu\right)&=-\frac{2T\Omega}{(1-e^2)^{1/3}}\rII^{-1} (t_1)=\frac{2 (e^2-2)\Omega}{e^4}t_1- \frac{4\sqrt{1-e^2}\Omega}{e^4}t_2\\
\rII^{-1}\left(\ad^*_{t_4}\mu\right)&=\frac{2T\Omega}{(1-e^2)^{1/3}}\rII^{-1} (t_2)=\frac{4\sqrt{1-e^2}\Omega}{e^4} t_1-\frac{2 (e^2-2)\Omega}{e^4}t_2
\end{align*}
From these expressions it
follows easily that
\begin{align*}
\Lambda (F,\mu) (t_1)&= -\frac{e^4+4e^2-8}{2e^4} \Omega t_3+\frac{4\sqrt{1-e^2}}{e^4}\Omega t_4\\
\Lambda (F,\mu) (t_2)&= \frac{-4\sqrt{1-e^2}}{e^4}\Omega t_3 + \frac{e^4+4e^2-8}{2e^4} \Omega t_4\\
\Lambda (F,\mu) (t_3)&= \frac{e^4+4e^2-8}{2e^4} \Omega t_1-\frac{4\sqrt{1-e^2}}{e^4}\Omega t_2\\
\Lambda (F,\mu) (t_4)&= \frac{4\sqrt{1-e^2}}{e^4}\Omega t_1 -\frac{e^4+4e^2-8}{2e^4} \Omega t_2
\end{align*}
 Finally, the entries of the
Arnold matrix are given by
$$\mathrm{Ar} (t_i,t_j) = \langle\mu, \ad_{t_i} \Lambda (F,\mu) (t_j)\rangle = \frac{2\sqrt{2}T\Omega}{(1-e^2)^{1/3}} \langle p, \ad_{t_i}
\Lambda (F,\mu) (t_j)\rangle ,\quad i=1,2,3,4.$$ Using
\eqref{severaladjoint2} the Arnold matrix is then given by
\begin{equation*}\mathrm{Ar}=\begin{bmatrix}
A_1 & -A_2 & 0   & 0\\
-A_2 & A_1 & 0   & 0\\
0   & 0   & A_1 & -A_2\\
0   & 0   & -A_2 & A_1\end{bmatrix}\quad\text{with}\quad \left\{\begin{matrix} A_1=\frac{(8-e^4 -4e^2)T\Omega^2}{e^4(1-e^2)^{\frac 13}}\\ \\
A_2=\frac{8(1-e^2)^{\frac 16} T\Omega^2}{e^4}
\end{matrix}\right.\end{equation*}

The trace and the determinant of each block of $\mathrm{Ar}$ are
positive and so $\mathrm{Ar}$ is positive definite.
\end{proof}

Next theorem gives the stability of the MacLaurin spheroids.
\begin{thm}\label{MacLaurin stability} A MacLaurin spheroid with eccentricity $e$  and momentum $\mu$ is:
\begin{itemize}
\item[(i)] nonlinearly $G_\mu$-stable if $e<e_0$ with $e_0 \simeq 0.952887$.
\item[(ii)] unstable if $e>e_0$.
\end{itemize}
\end{thm}
\begin{proof} For (i): As the Arnold form is non-degenerate, $G_\mu$-stability will follow whenever
$\left(\ed^2_FV_{\xi^\perp}+
\corr_{\xi^\perp}(F)\right)\restr{\Sigma_{\text{int}}}$ is positive
definite. For, recall from \eqref{sigmaint} that
$$\Sigma_{\text{int}} =
 \left\{\gamma_{\mathrm{SL}(3)}(F)+A : \,\, \gamma\in\mathfrak{q}^\mu,\, A\in\Sl^{\SL} ,\, \,\text{and}\,\, \left(\D\II(F) \cdot \left(\gamma_{\mathrm{SL}(3)}(F)+A\right)\right) (\xi^\perp)\in\mathfrak{p}^*\right\},
$$
where $\Sl^{\SL}$ is the linear slice for the $G$-action on
$\mathrm{SL}(3)$ at the oblate configuration, $F=\diag(a,a,c)$.
Matrices $A\in\Sl^{\SL}$ must verify \begin{eqnarray*}0  & = &\tr
[A^T(\widehat{\xi}F-F\widehat{\eta})],\quad\forall
\xi,\eta\in\R^3,\quad\text{and}\\ 0 & = &
\tr(F^{-1}A)\end{eqnarray*} because $A$ must belong, respectively to
the orthogonal complement to the tangent space to the group orbit
through $F$ and $A\in T_F\SL$. These two conditions give that $A$
must be of the form
$$A=\begin{bmatrix}
a_{11}&a_{12}&0\\
a_{12}&a_{22}&0\\
0&0&a_{33}
\end{bmatrix} \quad\text{with}\quad\frac{1}{a}(a_{11}+a_{22})+\frac{1}{c} a_{33}=0.
$$

 Therefore we can describe $\Sl^{\SL}$ as  the set  of  matrices of
the form
\begin{equation}\label{coordinate slice}A=\begin{bmatrix}
a_1+a_2 & a_3 & 0\\
a_3 & a_1-a_2 & 0\\
0 & 0 & -2\frac{c}{a} a_1
\end{bmatrix}
\end{equation}
with $a_1,a_2,a_3\in\R$. Let the vector $\gamma\in
\mathfrak{q}^\mu=\mathfrak{t}$ be
$\gamma=(\gamma_1,\gamma_2,\gamma_3,\gamma_4)$ with respect to the
basis $(t_1,t_2,t_3,t_4)$. Therefore, using \eqref{infgene}, we have
$$\gamma_{\mathrm{SL}(3)}(F)+A = \begin{bmatrix}
a_1+a_2&a_3&c \gamma_3-a\gamma_4\\
a_3& a_1-a_2&-c \gamma_1+a\gamma_2\\
-a \gamma_3+c \gamma_4& a \gamma_1- c\gamma_2&-2\frac{c}{a} a_1
\end{bmatrix}.
$$
The set $\Sigma_{\text{int}}$ is precisely the set of matrices
$\lambda_{\mathrm{SL}(3)}(F)+A$ for which $$\left(\D\II(F) \cdot
\left(\gamma_{\mathrm{SL}(3)}(F)+A\right)\right)
(\xi_L,\xi_R)^\perp\in\mathfrak{p}^*=
(\mathfrak{g}_F+\mathfrak{t})^\circ,$$ where
$(\mathfrak{h}+\mathfrak{t})^\circ$ denotes the annihilator of
$\mathfrak{h}+\mathfrak{t}$. Using $(\xi_L,\xi_R)^\perp =
\frac{\Omega}{\sqrt{2}} p$
 and $w\in\{h,t_1,t_2,t_3,t_4\}$
and  differentiating \eqref{lockedso},  the computation of
$$\left\langle w,\left(\D\II(F) \cdot
\left(\lambda_{\mathrm{SL}(3)}(F)+A\right)\right)
(\xi_L,\xi_+R)^\perp\right\rangle$$ gives, in terms of the
eccentricity $e$:
\begin{align*}
\langle h,\left(\D\II (F) \cdot \left(\gamma_{\mathrm{SL}(3)}(F)+A\right)\right) (\xi_L,\xi_R)^\perp\rangle&=0\\
 \langle t_1,\left(\D\II (F) \cdot \left(\gamma_{\mathrm{SL}(3)}(F)+A\right)\right) (\xi_L,\xi_R)^\perp\rangle&= \frac{T\Omega}{2 (1-e^2)^{1/3}} \left[(2+e^2) \gamma_3-2\sqrt{1-e^2} \gamma_4\right]\\
\langle t_2,\left(\D\II (F) \cdot \left(\gamma_{\mathrm{SL}(3)}(F)+A\right)\right) (\xi_L,\xi_R)^\perp\rangle&= \frac{T\Omega}{2 (1-e^2)^{1/3}} \left[2\sqrt{1-e^2} \gamma_3- (2+e^2) \gamma_4\right]\\
\langle t_3,\left(\D\II (F) \cdot \left(\gamma_{\mathrm{SL}(3)}(F)+A\right)\right) (\xi_L,\xi_R)^\perp\rangle&= \frac{T\Omega}{2 (1-e^2)^{1/3}} \left[-(2+e^2) \gamma_1+2\sqrt{1-e^2} \gamma_2\right]\\
\langle t_4,\left(\D\II (F) \cdot \left(\gamma_{\mathrm{SL}(3)}(F)+A\right)\right) (\xi_L,\xi_R)^\perp\rangle&= \frac{T\Omega}{2 (1-e^2)^{1/3}} \left[-2\sqrt{1-e^2} \gamma_1+ (2+e^2) \gamma_2\right]
\end{align*}

 It follows from the above expressions that $$\left(\D\II(F) \cdot
\left(\lambda_{\mathrm{SL}(3)}(F)+A\right)\right)
(\xi_L,\xi_R)^\perp\in (\mathfrak{g}_F+\mathfrak{t})^\circ$$ if and
only if $\gamma = 0$. This is equivalent to
$\Sigma_{\text{int}}=\Sl^{\SL}$.  Let us now compute the correction
term restricted to $\Sigma_{\text{int}}=\Sl^{\SL}$. For any
$A\in\Sl^{\SL}$ one has $\left(\D\II(F)\cdot A\right)
(\xi_L,\xi_R)^\perp =\frac{4T\Omega \sqrt{2}}{(1-e^2)^{1/6}} a_1 p$
and so
$$\Proj_{\mathfrak{t}^*\oplus\mathfrak{p}^*}\left[\left(\D\II(F)\cdot A\right) (\xi_L,\xi_R)^\perp\right] =
\left(\frac{4T\Omega \sqrt{2}}{(1-e^2)^{1/6}} a_1,0,0,0,0\right).
$$
From the   expression of $ \rII$  it is straightforward to obtain
$$\rII^{-1}\left(\Proj_{\mathfrak{t}^*\oplus\mathfrak{p}^*}\left[\left(\D\II(F)\cdot B\right) (\xi_L,\xi_R)^\perp\right]\right)=
\sqrt{2} \Omega (1-e^2)^{1/6} b_1 p,$$
where $b_1$ is the entry of $B\in \Sl^{\SL}$ playing the same role
of $a_1$ in $A$. Then, from \eqref{corrterm}
\begin{align*}
\corr_{(\xi_L,\xi_R)^\perp} (F) (A, B)&=\left\langle \frac{4T\Omega
\sqrt{2}}{(1-e^2)^{1/6}} a_1 p\, ,\,  \sqrt{2} \Omega (1-e^2)^{1/6}
b_1 p\right\rangle = 8T \Omega^2 a_1 b_1.
\end{align*}

The computation of $\ed^2_FV^\lambda_{(\xi_L,\xi_R)^\perp} (A,B)$ is
lengthy but with no difficulties. Using Lemma \ref{second derivative
augmented} we obtain

 \begin{align*}
\langle(\xi_L,\xi_R)^\perp, \left(\D^2_F\II (A,B)\right)(\xi_L,\xi_R)^\perp\rangle&= 4 a_1 b_1T \Omega^2\\
\ed^2_F \det\, (A, B) &=  -2  c (3 a_1b_1 +a_2 b_2 + a_3 b_3)
 \end{align*}
  We fix a basis for the slice $\Sl^{\SL}$ in which the coordinates of $A$ in \eqref{coordinate slice}  are $A= (a_1,a_2,a_3)$.
  With respect to this basis the matrix for $\ed^2_FV\rrestr{\Sigma_\mathrm{int}}$ is given by $\ed^2_FV\rrestr{\Sigma_\mathrm{int}}=\diag(D_1,D_2,D_2)$ with
\begin{align*}
\frac{a^{10}}{4} D_1=&a^4(a^6+2)V_1+ 3 a^6(a^6-1)V_2+ 4 (a^6-1)^2 V_{11}+8a^2(a^6-1)^2 V_{12}\\
& + 4 a^4(a^6-1)^2V_{22},\\
\frac{a^{10}}{4}D_2=&  a^{10}V_1+ a^6 (1-a^6)V_2.
\end{align*}

Therefore, with respect to this basis we have
\begin{equation}\label{stabformula}(\ed^2_FV_{\xi^\perp}^\lambda+\mathrm{corr}_{\xi^\perp}(F))\rrestr{\Sigma_\mathrm{int}}=
\diag(D_1+6c\lambda+6T\Omega^2\, , D_2+2c\lambda, D_2+2c\lambda) = \diag (S_1,S_2,S_2).
\end{equation}

For the MacLaurin spheroid we have, from Theorem \ref{theorem existence}, $\lambda = 2(c^2 V_1+ 2 c V_2)$  and $\Omega^2 = \frac{2}{T} e^2 (V_1+ a^2V_2)$. Then, in terms of the eccentricity, we have
\begin{align*}
S_1&=\frac{8}{1-e^2}\left((3-4e^2+e^4)V_1+3(1-e^2)^{2/3}V_2+2e^4(1-e^2)^{2/3}V_{11}\right.\\
&\left.+ 4e^4(1-e^2)^{1/3}V_{12}+2e^4V_{22}\right)\\
S_2&=\frac{4}{1-e^2}\left( (2-3e^2+e^4)V_1+(2-3e^2)(1-e^2)^{2/3}V_2\right)
\end{align*}

Expressing the partial derivatives of $V$ in terms of the integrals $J_O(k,r)$ we obtain
\begin{align*}
S_1&= \frac{2R}{e^5}\left(9e(3-5e^2+2e^4)-\sqrt{1-e^2}(27-36e^2+8e^4)\arcsin e\right)\\
S_2&= \frac{R}{e^5}\left((1-e^2)e(3+4e^2)-\sqrt{1-e^2}(3+2e^2-4e^4)\arcsin e\right)
\end{align*}

The plots of $S_1$ and $S_2$ are shown in Figure~\ref{plots
Maclaurin}. They show that $S_1$ is always positive in $(0,1)$ while
$S_2$ has a root $e_0\in (0,1)$  being positive for $e<e_0$ and
negative for $e>e_0$. We used the Mathematica programing system for
the numerical computation of $e_0$ to obtain $e_0\simeq 0.952887$.
Therefore MacLaurin spheroids are $G_\mu$-stable for $e<e_0$.

             \begin{figure}[h]
        \begin{center}
          \psfrag{0.2}{$0.2$}
          \psfrag{0.4}{$0.4$}
          \psfrag{0.6}{$0.6$}
          \psfrag{0.8}{$0.8$}
          \psfrag{1.0}{$1.0$}
          \psfrag{0.5}{$0.5$}
          \psfrag{1.5}{$1.5$}
          \psfrag{2.5}{$2.5$}
                    \psfrag{2.0}{$2.0$}
                                        \psfrag{3.0}{$3.0$}
  \resizebox{!}{3.5cm}{\includegraphics{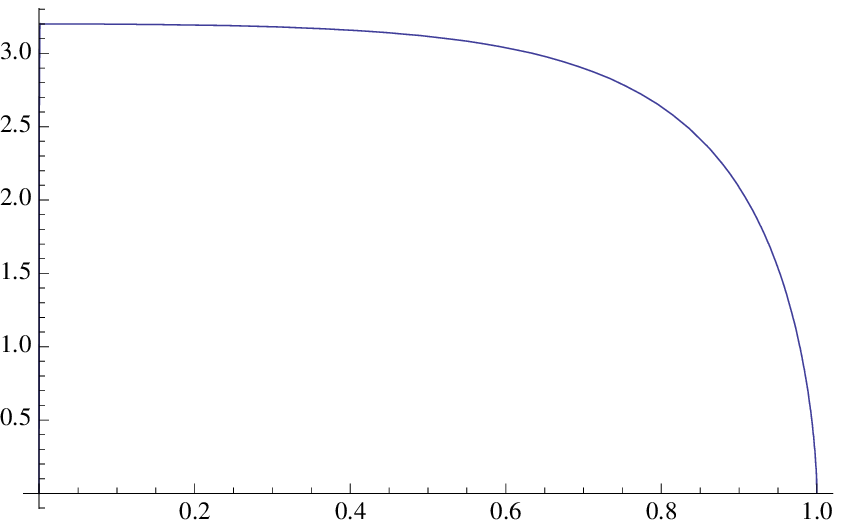}}  \hspace{0.3cm}    \resizebox{!}{3.5cm}{\includegraphics{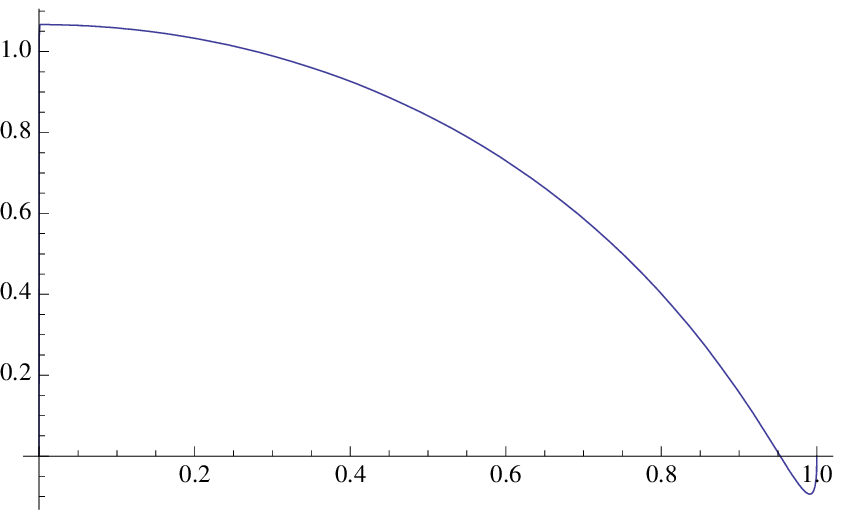}}
           \caption{The plots of $S_1$ (left) and $S_2$  (right)  in $R$ units.}
        \label{plots Maclaurin}
        \end{center}
    \end{figure}
For (ii): When  $e\in [e_0,1)$ we study the spectrum of the linearized Hamiltonian vector field $L_h=\w_N^{-1}(\ed_z^2 h_{\xi^\perp}\rrestr{N})$,
where in our case
  $N$ and $\w_N$ are as in Theorem \ref{symplectic mat}. Also, from Proposition 6.2 in \cite{miguelstab} we have

$$\left(\ed_F^2 h_{\xi^\perp}\right)\rrestr{N}=\begin{bmatrix}
\mathrm{Ar} & 0 & 0\\
 0 &  (\ed^2_FV^\lambda_{(\xi_L,\xi_R)^\perp}+\mathrm{corr}_{(\xi_L,\xi_R)^\perp})\rrestr{\Sl^{\SL}} & 0\\
 0 & 0 & \ll\cdot,\cdot\gg_{\Sl^{\SL*}}\end{bmatrix}
 $$
 Let us start by computing the block $\ll\cdot,\cdot\gg_{\Sl^{\SL*}}$ of the above matrix,
 where $\ll\cdot,\cdot\gg_{\Sl^{\SL*}}$ is the inner product on the dual of $\Sl^{\SL}$  induced by the Riemannian metric on $\Sl^{\SL}$.
 For, let as before fix the ordered basis $\{s_1,s_2,s_3\}$ on the slice
 $\Sl^{\SL}$ where
 $$s_1=\begin{bmatrix}
 1&0&0\\
 0&1&0\\
 0&0&-2 \sqrt{1-e^2}
 \end{bmatrix}\qquad s_2=\begin{bmatrix}
 1&0&0\\
 0&-1&0\\
 0&0&0
 \end{bmatrix}\qquad s_3=\begin{bmatrix}
 0&1&0\\
 1&0&0\\
 0&0&0
 \end{bmatrix}.
 $$
 Recalling that $\ll A,B\gg_{\Sl^{\SL}} =T \operatorname{tr} (A^TB)$ then the matrix that represents $\ll\cdot,\cdot\gg_{\Sl^{\SL}}$
 in the fixed basis  is
 $$\mathcal{R}_1=2T\begin{bmatrix} 3-2e^2 & 0 & 0\\
0 & 1 & 0\\
0 & 0 & 1\end{bmatrix}.$$

Let $\{s_1^*,s_2^*,s_3^*\}$ be the dual basis of $\{s_1,s_2,s_3\}$
under the identification of ${\Sl^{\SL*}}$ with ${\Sl^{\SL}}$ using
the inner product $ \ll \cdot,\cdot\gg_{\Sl^{\SL}}$. In this basis
the induced inner product $\ll\cdot,\cdot\gg_{\Sl^{\SL*}}$ is
represented by $\mathcal{R}_1^{-1}$.

 Let  $\mathcal{R}_2=(\ed^2_FV_{(\xi_L,\xi_R)^\perp}^\lambda+\mathrm{corr}_{(\xi_L,\xi_R)^\perp}(F))\rrestr{\Sigma_\mathrm{int}}$, then
$$ \left(\ed_F^2 h_{(\xi_L,\xi_R)^\perp}\right)\rrestr{N}=\begin{bmatrix}
\mathrm{Ar} & 0 & 0\\
0& \mathcal{R}_2&0\\
0&0&\mathcal{R}_1^{-1}
\end{bmatrix}.
$$

To compute $L_h$ in the basis $\{t_1,t_2,t_3,t_4,s_1,s_2,s_3,s_1^*,s_2^*,s_3^*\}$ for $N$ we use the formula for
$\w_N$ given in  Theorem~\ref{symplectic mat}.   Let us now compute each of the blocks of $\w_N$.

Recall that $\q^\mu=\mathfrak{t}$ for a MacLaurin spheroid. Then,
from Theorem \ref{symplectic mat}, for
$\gamma_1,\gamma_2\in\mathfrak{t}$ we have:
\begin{align*}
\Xi(\gamma_1,\gamma_2) &= -\mu\cdot \ad_{\gamma_1}\gamma_2 = -\left\langle\frac{2\sqrt{2}T\Omega}{(1-e^2)^{1/3}} p,\ad_{\gamma_1}\gamma_2\right\rangle.
\end{align*}
Using \eqref{severaladjoint2}, the matrix $\Xi$ is given by
$$\Xi=\frac{2T\Omega}{(1-e^2)^{1/3}}
  \begin{bmatrix}
  0&0&-1&0\\
  0&0&0&1\\
  1&0&0&0\\
  0&-1&0&0
 \end{bmatrix}
 $$
Since for a MacLaurin spheroid $\Sigma_{\text{int}}=\Sl^{\SL}$ it is
immediate from the definition of $\Psi$ in Theorem~\ref{symplectic mat}
 that $\Psi$ is the zero matrix.

 We now compute the Coriolis term $-\ed\chi^{(\xi_L,\xi_R)^\perp}\rrestr{\Sigma_\text{int}}$. For that we will obtain a concrete
 expression for the right hand side of the equality
\begin{equation}\label{cori0}
\ed\chi^{(\xi_L,\xi_R)^\perp}(X,Y)=X(\chi^{(\xi_L,\xi_R)^\perp}(Y))-Y(\chi^{(\xi_L,\xi_R)^\perp}(X))-\chi^{(\xi_L,\xi_R)^\perp}([X,Y]),\end{equation}
with $X,Y\in \mathfrak{X}(\mathrm{GL}^+(3))$.

Start by considering for $U,V\in T_I\mathrm{GL}^+(3)=\mathrm{L}(3)$
the corresponding left-invariant extensions
$X_U,X_V\in\mathfrak{X}(\mathrm{GL}^+(3))$. We have $X_U(F)=FU$ for
every $F\in\mathrm{GL}^+(3)$. Recall that
$(\xi_L,\xi_R)^\perp=\frac{\Omega}{2}(\mathbf{e_3},-\mathbf{e_3})$.
Then according to the definition given in Theorem \ref{symplectic
mat} we have
 \begin{equation}\label{cori1}\begin{array}{ll}
 \chi^{(\xi_L,\xi_R)^\perp}(X_U)(F) & =\frac{\Omega T}{2}\tr\left( (\widehat{\mathbf{e_3}}F+F\widehat{\mathbf{e_3}})^TFU\right)\vspace{3mm}\\ & =
 -\frac{\Omega T}{2}\tr\left(\widehat{\mathbf{e_3}}(F^TFU+FUF^T)\right).\end{array}\end{equation}

It is straightforward to obtain

\begin{equation}\label{cori2}\begin{array}{l}X_V(\chi^{(\xi_L,\xi_R)^\perp}(X_U))(F) =  \vspace{3mm}\\
=  -\frac{\Omega T}{2}\tr\left(\widehat{\mathbf{e_3}}((FV)^TFU+F^TFVU+FVUF^T+FU(FV)^T)\right)\vspace{3mm}\\
=   -\frac{\Omega T}{2}\tr\left(\widehat{\mathbf{e_3}}(V^TF^TFU+F^TFVU+FVUF^T+FUV^TF^T)\right)\end{array}\end{equation}

Also, since $X_U,X_V$ are left-invariant vector fields, the identity $[X_U,X_V]=X_{UV-VU}$ holds and we have, from  \eqref{cori1}
\begin{equation}\label{cori3}\chi^{(\xi_L,\xi_R)^\perp}([X_U,X_V])(F)=
-\frac{\Omega T}{2}\tr\left(\widehat{\mathbf{e_3}}(F^TF(UV-VU)+F(UV-VU)F^T)\right).\end{equation}

In order to compute $-\ed\chi^{(\xi_L,\xi_R)^\perp}\rrestr{\Sigma_\text{int}}$ let $F=\mathrm{diag}(a,a,c)$ and $A,B\in \Sigma_\text{int}$.
The unique left-invariant vector fields extending $A$ and $B$ are $X_{F^{-1}A}$ and $X_{F^{-1}B}$ respectively. Then, using \eqref{cori0} together with \eqref{cori2} and \eqref{cori3} it is immediate to obtain $-\ed\chi^{(\xi_L,\xi_R)^\perp}\rrestr{\Sigma_\text{int}}=0$.
 Therefore, from Theorem \ref{symplectic mat}  the symplectic matrix $\mathbf{\w}_N$ and its inverse are
  $$\mathbf{\w}_N =\begin{bmatrix}
 \Xi&0&0\\
 0&0&\mathbf{1}\\
 0&\mathbf{-1}&0
 \end{bmatrix}\qquad \mathbf{\w}^{-1}_N= \begin{bmatrix}
 \Xi^{-1}&0&0\\
 0&0&\mathbf{-1}\\
 0&\mathbf{1}&0
 \end{bmatrix}. $$
So the linearized vector field is
\begin{equation*}
L_h=\begin{bmatrix} \Xi^{-1}\mathrm{Ar} & 0 & 0\\
0 & 0 & -\mathcal{R}_1^{-1}\\
0 & \mathcal{R}_2 & 0\end{bmatrix},
\end{equation*}
where in our basis $\mathcal{R}_2= \operatorname{diag} (S_1,S_2,S_2)$ is given in \eqref{stabformula}.

The block $ \Xi^{-1}\mathrm{Ar}$ has imaginary eigenvalues $\epsilon_1^{\pm}=\pm i \frac{\sqrt{8+e^2}}{2e}\Omega$ with multiplicity 2.
For the block
$$\begin{bmatrix} 0 & -\mathcal{R}_1^{-1}\\
 \mathcal{R}_2 & 0\end{bmatrix}$$
 we obtain the following eigenvalues:
 \begin{itemize}
 \item $\epsilon_2^{\pm}=\pm i\sqrt{\frac{S_1}{(6-4e^2)T}}$ with multiplicity 1, and
 \item $\epsilon_3^{\pm}=\pm i\sqrt{\frac{S_2}{4T}}$ with multiplicity 2.
 \end{itemize}

 As $(6-4e^2)$ and $S_1$ are positive for $0<e<1$, then $\epsilon_2^{\pm}$ is always imaginary in that range. However,
 $\epsilon_3^{\pm}$ becomes real if $S_2$ becomes negative, which happens exactly at $e_0$ as seen from the previous
 stability analysis. Hence, if $e>e_0$ the MacLaurin spheroid becomes linearly unstable, therefore unstable. This loss
 of stability corresponds exactly to a collision at 0 of $\epsilon_3^{\pm}$ for $e=e_0$, which passes from being pure imaginary to be real.
\end{proof}

\begin{rem}\label{rem maclaurin}\

\noindent 1.  In Riemann's work \cite{Riemann} some conclusions were
made concerning the stability of  Maclaurin spheroids by studying
the existence of a minimum of a certain function $G$.  The existence
of this minimum was not done by studying the second variation of
$G$. Riemann even says in page 188: ``The direct analysis of the
second variation of $G$ when the first variation vanishes would be
very complicated; we can however decide if this function has a
minimum by the following form:...''. He follows with the analysis of
the behavior of $G$. His final conclusion on the stability ends the
paragraph 9 of his paper and is the following: ``From this study it
follows that the case of a rotation of an oblate ellipsoid, around
its shortest axis, case already known to MacLaurin, can only be
unstable if the relation between the shortest axis with the others
is less than 0.303327...". We note that if the relation  between the
shortest axis of the Maclaurin spheroid and the others is
$\frac{c}{a}< 0.303327$ this is equivalent to say that the
eccentricity $e>e_0=0.952887$.

The value $0.303327$ obtained by Riemann follows from his study on
the existence of oblate spheroids in pages 184--185 of
\cite{Riemann}, namely as the root of the last displayed equation in
page 184. We remark that this equation is equivalent to the equation
$S_2= 0$ where $S_2$ is as in the  proof of (i) in Theorem
\ref{MacLaurin stability}. Indeed
\begin{align*}
S_2=0 &\Longleftrightarrow e
(1-e^2)(3+4e^2)-\sqrt{1-e^2}(3+2e^2-4e^4)\arcsin e=0.
\end{align*}
Taking $e= \cos \psi = \sin(\frac{\pi}{2}-\psi)$ and $\psi\in (0, \frac{\pi}{2})$ one has
\begin{align}
S_2=0&\Longleftrightarrow \nonumber\\
&\hspace{-0.4cm}\cos \psi \sin^2\psi \left(5+ 2\cos (2\psi) \right)-\sin\psi \left(\frac{5}{2}-\cos (2\psi)-\frac{1}{2} \cos (4\psi)\right) (\frac{\pi}{2} -\psi)=0\nonumber\\
&\hspace{-0.6cm}\Longleftrightarrow 10 \sin (2\psi)+2 \sin (4\psi)+ \left(-5+ 2\cos (2\psi)+ \cos (4\psi)\right)  \left(\pi-2 \psi\right)=0. \label{Riemann184}
\end{align}
The equation~\eqref{Riemann184} is the same one appearing  in
Riemann's paper.  With respect to the computation of the root of
this equation Riemann just says: ``...this equation has, for $\psi$
between $0$ and $\pi/2$, the unique root $\sin \psi =
0.303327$...''.  Indeed, this is equivalent to say that $e_0 = \cos
\psi = 0.952887...$

\noindent 2. Concerning the stability of MacLaurin spheroids, it is
shown in Liapunov's work \cite{Liapunov}  that  under the hypothesis
of the preservation of the ellipsoidal shape (setup we used)  the
value $e_0$ of loss of stability is the same as in Riemann's work,
but if this hypothesis is dropped then the MacLaurin spheroid is
only stable for $e<e_1$ with $e_1= 0.8126...$. The point $e_1$   is
exactly the point where the family of MacLaurin spheroids bifurcates into a branch of 
ellipsoids with 3 distinct axes lengths (Jacobi ellipsoids). We refer
the reader to pages 52 and 61-63 of  \cite{Liapunov}.
\end{rem}

                 \subsection{Transversal spheroids}

All the qualitative properties, including stability, of two relative
equilibria lying in the same orbit of the symmetry group are the
same. Therefore, in view of Remark \ref{transversal adjoint}, in
this subsection we analyze the nonlinear stability of the $+$ family
of transversal spheroids and the main result, Theorem
\ref{transversal stability}, will follow for both families. We will
set $\w_+=\w$, $f=f_+$ and $(\xi_L,\xi_R)_+=(\xi_L,\xi_R)$ for
notational simplicity. Also, to keep the notation consistent with the proof of Theorem~\ref{theorem existence}, we will set $\mathbf{n}=\mathbf{e_2}$.

 In this case the computation of the splitting \eqref{split} is simplified due to the fact that $\g_{p_x}=\{0\}$ and therefore
 $\mathfrak{p}=\mathfrak{g}_\mu$. Introducing the vectors
 $h=\frac{1}{\sqrt{2}}(\mathbf{e_3},\mathbf{e_3}),\, p_1=(\mathbf{e_2},0),\,p_2=(0,\mathbf{e_2}),\,t_1=
 (\mathbf{e_1},0),\,t_2=(0,\mathbf{e_1}),\,t_3=\frac{1}{\sqrt{2}}(\mathbf{e_3},-\mathbf{e_3})$, we choose
 $$\begin{array}{cl}
 \g_F  & =  \mathrm{span}\{h\}\\
 \mathfrak{p} &  =  \mathrm{span}\{p_1,p_2\}\\
 \mathfrak{t}  & =  \mathrm{span}\{t_1,t_2,t_3\}.
 \end{array}$$
 These subspaces are obviously invariant under the action of $G_{P_F}=\mathbb{Z}_2(\mathbf{e_2})$.
With respect to the basis $(p_1,p_2,t_1,t_2,t_3)$ for $\mathfrak{p}\oplus\mathfrak{t}$ we have,
$$\rII = T\left[\begin{array}{ccccc}
a^2+c^2 & -2ac &0 & 0 & 0\\
-2ac & a^2 + c^2 & 0 & 0 & 0 \\
0 & 0 & a^2 + c^2 & -2ac & 0\\
0 & 0 & -2ac & a^2+c^2  \\
0 & 0 & 0 & 0 & 4a^2\end{array}
    \right].$$
    Or in terms of the eccentricity of a prolate spheroid,
    $$\rII = T\left[\begin{array}{ccccc}
\frac{2-e^2}{(1-e^2)^{2/3}} & \frac{-2}{(1-e^2)^{1/6}} &0 & 0 & 0\\
\frac{-2}{(1-e^2)^{1/6}} & \frac{2-e^2}{(1-e^2)^{2/3}} & 0 & 0 & 0 \\
0 & 0 & \frac{2-e^2}{(1-e^2)^{2/3}} & \frac{-2}{(1-e^2)^{1/6}}& 0\\
0 & 0 & \frac{-2}{(1-e^2)^{1/6}} & \frac{2-e^2}{(1-e^2)^{2/3}} & 0  \\
0 & 0 & 0 & 0 & 4(1-e^2)^{1/3} \end{array}
    \right].$$

It follows immediately that $\mathfrak{p}$ and $\mathfrak{t}$ are
orthogonal with respect to $\rII$, so our choice of the splitting
$\g=\g_F\oplus\mathfrak{p}\oplus\mathfrak{t}$ is correct. In this
basis the orthogonal velocity is
$$(\xi_L,\xi_R)^\perp =\w (p_1+f \,p_2),$$
with $\w$ and $f$ as in the $+$ family in Theorem \ref{theorem
existence}.

It is straightforward to compute the adjoint representation of $\g$ in this basis:
\begin{lemma}\label{transadjoint}
The elements of $\mathfrak{p}\oplus\mathfrak{t}$ satisfy the following relations:
\begin{eqnarray*}
 \ad_{t_1}t_2  = 0 & \quad \ad_{t_1}t_3=-\frac{1}{\sqrt{2}}p_1\quad\ad_{t_2}t_3=\frac{1}{\sqrt{2}}p_2\\
\ad_{t_1}p_1  =\frac{1}{\sqrt{2}} (h+t_3) & \quad \ad_{t_1}p_2=0\\\
\ad_{t_2}p_1=0 & \quad\ad_{t_2}p_2=\frac{1}{\sqrt{2}}(h-t_3)\\
\ad_{t_3}p_1=-\frac{1}{\sqrt{2}}t_1& \quad \ad_{t_3}p_2=\frac{1}{\sqrt{2}}t_2\\
\ad_{p_1}  p_2=0. &
\end{eqnarray*}
\end{lemma}
With this, we can prove the following proposition, which shows that the stability method is applicable.
\begin{prop}\label{arnoldtransversal}
The Arnold form for a transversal spheroid is positive definite.
\end{prop}
\begin{proof}
Recall that the momentum of a transversal spheroid is
$$\mu =\rII(\xi_L,\xi_R)^\perp =T\w \left(\kappa^L  p_1+\kappa^R  p_2\right) ,$$
where $\kappa^L :=\left[(a^2+c^2)-(2/a)f \right]$ and $\kappa^R
:=\left[(a^2+c^2)f -(2/a)\right]$. From Lemma \ref{transadjoint},
and recalling that $\ad^*_\gamma\rho=-\ad_\gamma\rho$, we have, for
$\gamma=\gamma^{(1)}t_1+\gamma^{(2)}t_2+\gamma^{(3)}t_3\in\mathfrak{t}$:
$$\Proj_{\g_F}\left(\ad^*_\gamma\mu\right)=\frac{-T\w }{\sqrt{2}}\left(\gamma^{(1)}\kappa^L +\gamma^{(2)}\kappa^R \right)h.$$
This is zero iff $\gamma^{(2)}=-\kappa \gamma^{(1)}$, with
$$\kappa :=\frac{\kappa^L }{\kappa^R }=-\frac{(e-1)(e+ 2)}{(e- 2)\sqrt{1-e^2}},$$
and therefore $\q^\mu=\{\gamma^{(1)}(t_1-\kappa
t_2)+\gamma^{(3)}t_3\, :\, (\gamma^{(1)},\gamma^{(3)})\in\R^2\}$. A
basis for $\q^\mu$ is given by $\{\gamma_a=t_1-\kappa
t_2,\gamma_b=t_3\}$. Now, proceeding as for the MacLaurin spheroid,
we compute
\begin{eqnarray*}
\ad_{\gamma_a}\left(\rII^{-1}\mu \right)=\frac{\w
}{\sqrt{2}}\left((1-\kappa  f )h+(1+\kappa  f )t_3 \right) &
\ad_{\gamma_b}\left(\rII^{-1}\mu \right)=\frac{\w }{\sqrt{2}}(f  t_2-t_1)\\
\ad^*_{\gamma_a}\mu = -\sqrt{2}T\w \kappa^L  t_3 &  \ad^*_{\gamma_b}\mu = \frac{T\w }{\sqrt{2}}(\kappa^L  t_1-\kappa^R  t_2)\\
\rII^{-1}(\ad^*_{\gamma_a}\mu)= \frac{-\kappa ^L\w }{2\sqrt{2}(1-e^2)^{1/3}}t_3    & \\
\end{eqnarray*}
\begin{eqnarray*}\rII^{-1}(\ad^*_{\gamma_b}\mu)=  & \frac{(1-e^2)^{1/6}\w }{\sqrt{2}e^4}\left( (-\sqrt{1-e^2}(e^2-2)\kappa ^L+2(e^2-1)\kappa ^R)t_1\right. \\ & \left. +
(\sqrt{1-e^2}(e^2-2)\kappa ^R-2(e^2-1)\kappa ^L)t_2 \right).\end{eqnarray*}

From where it easily follows that, in the basis $\{\gamma_a,\gamma_b\}$ for $\q^\mu$,
$$\mathrm{Ar} =\mathrm{diag}\left(\frac{3e^4(2+ e)T\w ^2}{2(2- e)(1-e^2)^{5/3}},\frac{4(1+ e)(2-e)(2+e)T\w ^2}{e^2(1-e^2)^{2/3}}\right).$$
The entries of $\mathrm{Ar}$ are obviously positive.
\end{proof}

We can therefore apply the singular reduced energy-momentum method to study the stability of the transversal spheroid, obtaining the following
\begin{thm}\label{transversal stability}
Both families of transversal spheroids are nonlinearly stable for
all eccentricities.
\end{thm}

\begin{figure}[h]
        \begin{center}
          \psfrag{0.2}{$0.2$}
          \psfrag{0.4}{$0.4$}
          \psfrag{0.6}{$0.6$}
          \psfrag{0.8}{$0.8$}
          \psfrag{1.0}{$1.0$}
          \psfrag{0.5}{$0.5$}
          \psfrag{1.5}{$1.5$}
          \psfrag{2.5}{$2.5$}
                    \psfrag{2.0}{$2.0$}
                                        \psfrag{3.0}{$3.0$}
  \resizebox{!}{3.5cm}{\includegraphics{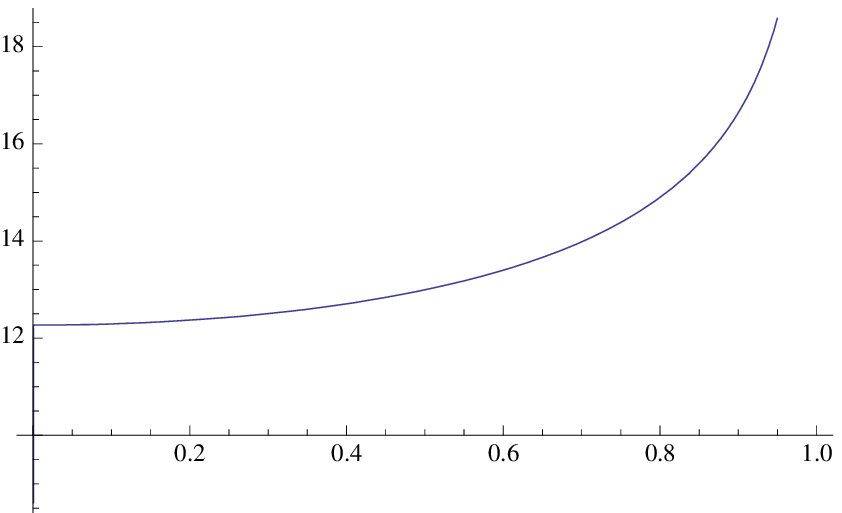}}  \hspace{0.3cm}    \resizebox{!}{3.5cm}{\includegraphics{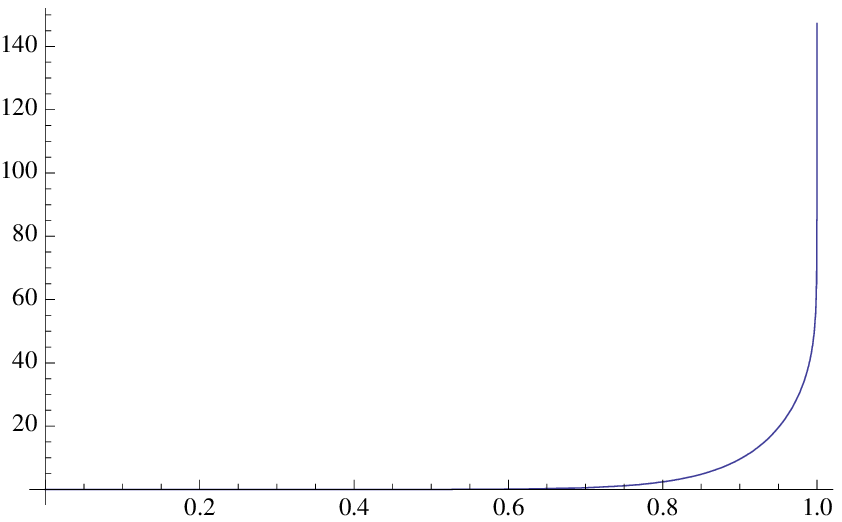}}
           \caption{The plots of $\operatorname{tr}\,(U)$ (left) and $e^{10} \det (U)$  (right),  in $R$ and $R^2$ units respectively.}
        \label{prolateplots}
        \end{center}
    \end{figure}

\begin{proof}
We start by computing the space of internal variations $\Sigma_\text{int}$. Recall that the slice $\Sl^{\SL}$ at
$F=\text{diag}(a,a,c)$ is given by \eqref{coordinate slice}.  Hence, for $\gamma=(\gamma^{(1)},\gamma^{(3)})$ in $\q^\mu$ and $A=(a_1,a_2,a_3)$ in $\Sl^{\SL}$ we have

$$\gamma_{\SL}(F)+A=\left[\begin{array}{ccc}
a_1+a_2 & a_3-\sqrt{2}a\gamma^{(3)} & 0\\
a_3+\sqrt{2}a\gamma^{(3)}  & a_1-a_2 & -(c+a\kappa )\gamma^{(1)}\\
0 & (a+c\kappa )\gamma^{(1)} & \frac{-2c}{a}a_1\end{array}
\right].$$

Differentiating \eqref{lockedso} we find
$$\begin{array}{ll}
\langle h,\left(\D\II (F) \cdot \left(\gamma_{\mathrm{SL}(3)}(F)+A\right)\right) (\xi_L,\xi_R)^\perp\rangle & = 0\\

\langle t_1,\left(\D\II (F) \cdot \left(\gamma_{\mathrm{SL}(3)}(F)+A\right)\right) (\xi_L,\xi_R)^\perp\rangle & = \frac{2(e+ 1)T\w
(ea_3+\sqrt{2}(1-e^2)^{1/6}\gamma^{(3)})}{(1-e^2)^{5/6}}
\\

\langle t_2,\left(\D\II (F) \cdot \left(\gamma_{\mathrm{SL}(3)}(F)+A\right)\right) (\xi_L,\xi_R)^\perp\rangle & = \frac{- 2T\w (ea_3
+\sqrt{2}(1-e^2)^{1/6}\gamma^{(3)})}{(1-e^2)^{1/3}}
\\

\langle t_3,\left(\D\II (F) \cdot \left(\gamma_{\mathrm{SL}(3)}(F)+A\right)\right) (\xi_L,\xi_R)^\perp\rangle & = \frac{3\sqrt{2}e^3T\w \gamma^{(1)}}{(e-
2)(1-e^2)^{2/3}}.
\end{array}$$
In order for $\left(\mathbf{D}\II(F)\cdot
(\gamma_{\Sl^{\SL}}+A)\right)(\xi_L,\xi_R) ^\perp\in
(\mathfrak{g}_F\oplus \mathfrak{t})^\circ$ all the above expressions
must vanish, which happens if and only if $\gamma^{(1)}=0$ and
$\gamma^{(3)}=a_3\epsilon $, with $\epsilon =\frac{-
e}{\sqrt{2}(1-e^2)^{1/6}} $. Therefore we can chose a basis for the
space of internal variations $\Sigma_\text{int}$ such that any
element $v$ belonging to it has components $(a_1,a_2,a_3)$ with the
parametrization
$$v= (a_1,a_2,a_3)\mapsto \left[\begin{array}{ccc}
a_1+a_2 & a_3(1-\sqrt{2}\,a\,\epsilon) & 0\\
a_3(1+\sqrt{2}\,a\,\epsilon)  & a_1-a_2 &0\\
0 & 0 & \frac{-2c}{a}a_1\end{array}  \right].$$
It is straightforward to obtain
$$\left(\mathbf{D}\II(F)\cdot  v \right)(\xi_L,\xi_R) ^\perp=2eT\w \left(
\frac{-\left( (e- 1)a_1+(e+
1)a_2\right)}{(1-e^2)^{5/6}}p_1+\frac{(e+
1)a_1+a_2(e-1)}{(e-1)(1-e^2)^{1/3}}p_2 \right) .$$ Then the correction term,  for $v_1=(a_1,a_2,a_3),\,v_2=(b_1,b_2,b_3)$, is given by:
$$\mathrm{corr}_{(\xi_L,\xi_R) ^\perp}(v_1,v_2)  =\frac{8 T\w ^2}{e- 1}\left(
\frac{9-5e^2}{e^2-1}a_1b_1-3(a_1b_2+a_2b_1)-a_2b_2
\right)
$$
For the restriction of the Hessian of $V^\lambda_{(\xi_L,\xi_R)
^\perp}$ at $F$ we compute first the following second variations

$$\ed_F^2V(v_1,v_2) = D_1a_1b_1+D_2a_2b_2+D_3a_3b_3,\quad\mathrm{with}$$
\begin{eqnarray*}
D_1 & = & \frac{4}{(1-e^2)^{5/3}}\left(-(1-e^2)^{2/3}(-3+e^2)V_1
+3e^2(e^2-1)V_2+4e^4V_{11}\right.\\
& & \left. +8e^4(1-e^2)^{1/3}V_{12}+4e^4(1-e^2)^{2/3}V_{22}
\right)\\
D_2 & = & 4\left( V_1+\frac{e^2}{(1-e^2)^{2/3}}V_2\right)\\
D_3 & = & \frac{4}{e^2-1}\left((e^4-1)V_1+e^2(1-e^2)^{1/3}(e^2-3)V_2
\right)
\end{eqnarray*}

$$\begin{array}{ll}
\langle (\xi_L,\xi_R)^\perp
,(\mathbf{D}_F^2\II(v_1,v_2))(\xi_L,\xi_R)^\perp \rangle =& \vspace{2mm}\\ 4T\w
^2\left(\frac{(9-5e^2)}{(e-1)^2(1+ e)}a_1b_1+\frac{3}{1- e}(a_1b_2+a_2b_1)
+\frac{1}{(1-
e)}a_2b_2+(1+ e)a_3b_3\right) & \end{array}$$
$$
\ed_F^2\mathrm{det}\,(v_1,v_2)  =  \frac{-2}{(1-e^2)^{1/3}}\left(
3a_1b_1+a_2b_2+(1-e^2)a_3b_3 \right)$$

Putting all the contributions together, and substituting the
integral expressions for the derivatives of the potential and the
values of $\lambda$ and $\w$ given in Theorem~\ref{theorem
existence}, we find
that the matrix representing
$\ed^2_FV_{(\xi_L,\xi_R)^\perp}^\lambda+\mathrm{corr}_{(\xi_L,\xi_R)^\perp}(F)\rrestr{\Sigma_\text{int}}$
in the fixed basis of $\Sigma_\text{int}$ is block diagonal, with a
$2\times 2$ symmetric block in the first two components, and the
other block given by the coefficient of $a_3 b_3$, say $\phi$, which
has the following expression:
$$\phi= \frac{2}{(1-e^2)^{2/3}}\left[ (1-e^2)^{2/3} (3+e^2) V_1+ (3+ 2 e^2-e^4) V_2\right].$$
It is clear that $\phi$ is  positive since  $ (3+ 2 e^2-e^4)$ is
positive in $(0,1)$ and $V_1$ and $V_2$ are always positive.

The $2\times 2$ block is $U=\frac{R}{e^5}\begin{bmatrix}
x&y\\
y&z
\end{bmatrix}$ with
\begin{align*}
x&= - 27 e (1-e^2)+  (27-36 e^2 +17 e^4)  \mathrm{arctanh}\, e\\
y&= - 9 (1-e^2) \left[3 e - (3-e^2) \mathrm{arctanh}\, e \right]\\
z&= -2 e (3-4 e^2)+ 2 (3-5 e^2 + 2 e^4) \mathrm{arctanh}\, e
\end{align*}

The matrix $U$ has real eigenvalues.  The trace and the determinant
of $U$ are:
$$\operatorname{tr}\, (U) = \frac{ R}{e^5} \left[ - e (33- 35 e^2) + (33 - 46 e^2+ 21 e^4) \mathrm{arctanh}\, e \right]$$
\noindent and
\begin{align*}
\det (U)&=\frac{R^2}{e^{10}}\left[-27 e^2 (21- 40 e^2+ 19 e^4)\right.\\ & \left. -  2  (- 567 e+ 1269 e^3- 831 e^5+121 e^7) \mathrm{arctanh}\, e \right.\\
&-\left. (567-1458 e^2+ 1212 e^4-334 e^6+ 13 e^8) \mathrm{arctanh}^2
e\right].
\end{align*}
These two quantities are positive in the interval $e\in (0,1)$. Their plots  are displayed
in Figure~\ref{prolateplots}. Therefore, both families of transversal spheroids are
stable for all eccentricities.
\end{proof}


\begin{thebibliography}{99}


\bibitem{AMM81} Arms, J.M., Marsden, J.E., and Moncrief, V. [1975], Symmetry and bifurcations of momentum mappings. {\it Comm.
Math. Phys}, {\bf 78} 455--478.

\bibitem{Arnold2} Arnold, V.I. [1966], Sur la g\'eometrie differentielle
des groupes de Lie de dimension infinie et ses applications \`{a}
l'hydrodynamique des fluids parfaits, {\it Ann. Inst. Fourier,
Grenoble}, {\bf 16}, 319--361.

\bibitem{Chand} Chandrasekhar, S. [1987], {\it Ellipsoidal figures of equilibrium}, Dover Pub. Inc., New York.

\bibitem{Ciarlet}
Ciarlet P.G. [1993], {\em Mathematical Elasticity, volume 1: Three
dimensional elasticity},  North-Holland.


\bibitem{CoMu}  Cohen, H., and Muncaster, R.G. [1988], {\it The theory of pseudo-rigid bodies}, New York: Springer-Verlag.

\bibitem{FaDe} Fass\`o, F. and Lewis, D. [2001], Stability properties of Riemann ellipsoids, {\it Arch. Rational Mech. Anal.} {\bf 158}, 259--292.


\bibitem{LermanSinger} Lerman, E. and Singer, S.F.  [1998], Stability and persistence of relative equilibria
at singular values of the moment map, {\it Nonlinearity}, {\bf 11}, 1637-1649.

   \bibitem{DebraBif} Lewis, D.  Marsden, J., and Ratiu, T.S. [1987],
Stability and bifurcation of a rotating planar liquid drop, {\it J.
Math. Phys.}, {\bf 28}, 10, 2508--2515.

\bibitem{DeSim}  Lewis, D., Simo, J. C. [1990], Nonlinear stability of rotating pseudo-rigid bodies, {\it Proc. Roy. Soc. London Ser. A}, {\bf Vol. 427}, 1873, 281--319.


\bibitem{Liapunov} Liapounoff, M.A.  [1904], {\it Sur la stabilit\'e des figures ellipsoidales d'\'equilibrie d'un liquide
anim\'e d'un movement de rotation}, Annales de la Facult\'e des
Sciences de L'Universit\'e de Toulouse.

\bibitem{MarsdenLN} Marsden, J.E. [1992], {\it Lectures on Mechanics}, Lecture Note Series
{\bf 174}, LMS, Cambridge University Press.

\bibitem{MaHu}
 Marsden, J.E. and Hughes, T.J.R.  [1983], {\em Mathematical
Foundations of
 Elasticity}. Prentice-Hall.

\bibitem{Montaldi} Montaldi J. [1997], Persistence and stability of relative equilibria, {\it Nonlinearity}, {\bf 10}, 449-466.

\bibitem{OrtegaRatiu} Ortega, J-P. and Ratiu, T.S.  [1999], Stability of Hamiltonian relative equilibria, {\it Nonlinearity}, {\bf 12}, 693-720.

\bibitem{Patrick92} Patrick, G.W. [1992], Relative equilibria in Hamiltonian systems: the dynamics interpretation of nonlinear
stability on the reduced phase space, {\it J. Geom. Phys.}, {\bf 9}, 111-119.

\bibitem{PaRoWu} Patrick, G.W., Roberts, M. and Wulff, C.  [2004], Stability of Poisson  and Hamiltonian relative equilibria
by energy methods, {\it Arch. Rational Mech. Anal.}, {\bf 174}, 301-344.

\bibitem{PerlRodSou}  Perlmutter, M., Rodr\'{\i}guez-Olmos, M.  and Sousa-Dias, M.E. The symplectic normal space of a
cotangent-lifted action, {\it Differential Geometry and Applications, to appear}. Preprint math.SG/051207.

\bibitem{Poincare}  Poincar\'e, H. [1885], Sur L'Equilibre d'une Masse Fluide Anim\'ee d'un Mouvement de Rotation,
 {\it Acta Mathematica}, {\bf 7}, 259--380.

\bibitem{Riemann} Riemann, B. [1861],  Ein Beitrag zu den Untersuchungen \"{u}ber die Bewegung cines fl\"{u}ssigen gleichartigen Ellipsoides, {\it Abh. d. K\"{o}nigl. Gesell. der Wis. zu G\"{o}ttingen}, {\bf 9},  168--197.

\bibitem{MEres} Roberts, M. and Sousa-Dias, M.E. [1999],  Symmetries of Riemann ellipsoids, {\it Resenhas- IME. USP}, Vol. 4,  2, 183-221.

\bibitem{miguelstab} Rodr\'{\i}guez-Olmos, M. [2006], Stability of relative equilibria in simple mechanical systems,
{\it Nonlinearity}, {\bf 19}, no.4 , 853-877.

\bibitem{RSD} Rodr\'{\i}guez-Olmos, M. and  Sousa-Dias, M.E.  [2002], Symmetries of relative equilibria for simple mechanical systems.
 SPT 2002: Symmetry and perturbation theory (Cala Gonone),
 221--230, World Sci. Publ.

\bibitem{Ros88a} Rosensteel, G. [1988], Rapidly rotating nuclei as {R}iemann ellipsoids, {\it Ann. Physics}, {\bf 186}, 230--291.

\bibitem{Ros88b} Rosensteel, G. [1988], Geometric quantization of {R}iemann ellipsoids, {\it Group theoretical methods in physics (Varna, 1987)}. Lecture Notes in Phys, {\bf 313}, 253--260.

\bibitem{Ros01} Rosensteel, G. [2001], Gauge theory of {R}iemann ellipsoids, {\it J. Phys. A}, {\bf 34}, 13, L169--L178.

\bibitem{Simo} Simo J.C., Lewis, D. and Marsden J.E. [1991], Stability of relative equilibria. Part I: The reduced
energy-momentum method, {\it Arch. Rational Mech. Anal.}, {\bf 115}, 15--59.

\end{thebibliography}
\end{document}